\documentclass[12pt]{article}

\usepackage[letterpaper,margin=1in]{geometry}
\usepackage{setspace}
\usepackage{amsmath,amssymb,bm}
\usepackage{siunitx}
\usepackage{microtype}

\usepackage{graphicx}
\graphicspath{{figures/}}
\usepackage{booktabs}
\usepackage{array}
\usepackage{caption}
\usepackage[section]{placeins}
\usepackage{xcolor}

\usepackage[numbers,super,sort&compress]{natbib}
\usepackage[hidelinks]{hyperref}

\newcommand{\SV}{\ensuremath{S/V}}
\newcommand{\svin}{\ensuremath{\langle 4/d\rangle_{V}}}  

\begin{document}

\begin{titlepage}
\begin{flushleft}
{\LARGE\bfseries
Two observables of one wall: how surface relaxivity can bias the diffusion
intra-axonal fraction and the myelin water fraction\par}
\vspace{1.5em}

{\large Rutger H.\,J.\ Fick\textsuperscript{1,$\ast$}\par}
\vspace{1em}

\textsuperscript{1}\,dmrai-lab, Independent Research Initiative
(\url{https://dmrai-lab.org})\par
\vspace{2em}

\textbf{Corresponding author:}\\
\textsuperscript{$\ast$}\,Rutger H.\,J.\ Fick\\
dmrai-lab, Independent Research Initiative\\
\texttt{rutger.fick@dmrai-lab.org}\par
\vspace{2em}

\textbf{Running head:}\; Surface relaxivity biases microstructure fractions\par
\vspace{2em}

\textbf{Funding:}\; None.\par
\vspace{0.5em}
\textbf{Conflicts of interest:}\; The author declares no competing interests.\par
\end{flushleft}
\end{titlepage}

\begin{abstract}
\noindent
\textbf{Purpose:} Surface relaxivity and time-dependent diffusion are two readouts
of the same wall collisions on one substrate. The transverse rate microstructure imaging
fits is a bulk rate plus a surface rate $\rho\,(S/V)$; intra- and extra-axonal water carry
\emph{different} $S/V$, so their $T_2$ differ, biasing any compartment estimate that
normalises by a TE-weighted $b=0$. We quantify the resulting bias on the diffusion
intra-axonal signal fraction and on the myelin water fraction (MWF).

\textbf{Methods:} Closed forms for the interior (Brownstein--Tarr) and exterior
(Novikov--Burcaw) surface rates over myelinated cylinders are derived and validated by
wall-counting Monte Carlo. A calibre-distributed forward model, and grid-free closed forms,
give the intra-axonal fraction ($f_{\mathrm{intra}}$) bias vs packing and echo time and the
MWF bias vs calibre; a within-subject demyelination trajectory traces the longitudinal case.

\textbf{Results:} The interior/exterior $S/V$ ratio is $(1-\mathrm{VF})/(g\,\mathrm{VF})$,
independent of the calibre distribution, crossing unity at fibre volume fraction
$\mathrm{VF}^{\ast}=1/(1+g)$. Physiological white matter sits above it, so surface relaxivity
\emph{over}-weights the intra-axonal signal --- the leading-order intra-axonal fraction bias ---
by ${\approx}12\%$ over the robust packing band at clinical PGSE ($\mathrm{TE}=80$\,ms,
cited $\rho$). That fractions are TE-dependent is known; what is new is the surface-relaxivity
and packing \emph{attribution} and the closed-form sign law, with a testable
packing-dependent TE drift. The same physics reads through relaxometry as a smaller MWF bias:
the thinnest axons' water crosses below the myelin window and is counted as myelin, so fine WM
reads myelin-richer (${\sim}0.33$\,pp at cited $\rho$, far beneath single-voxel noise and
exposed only by regional averaging, but super-linear in the order-of-magnitude-uncertain $\rho$). In primary (lumen-preserving)
demyelination the MWF bias is lumen-set, nearly constant, and cancels in the change.

\textbf{Conclusion:} One $S/V$ surface-relaxivity physics biases both diffusion and
relaxometry microstructure estimates; the $f_{\mathrm{intra}}$ bias is first-order and
packing-dependent (a spatially structured systematic, scaling with the uncertain $\rho$),
the MWF bias small and structural.

\vspace{1em}
\end{abstract}

\section{Introduction}
\label{sec:intro}

White-matter microstructure imprints more than one MR contrast, and sometimes the
same physics drives two of them. When a water molecule diffuses among myelinated
axons it repeatedly strikes their walls, and each collision does two things at once:
it interrupts the molecule's displacement --- the event whose statistics,
accumulated over the fibre packing, make the extra-axonal diffusion coefficient
time-dependent \citep{Novikov2014, Burcaw2015} --- and it exposes the spin to the
wall's relaxation sink, adding \emph{surface relaxivity} \citep{Brownstein1979} at a
rate proportional to the pore surface-to-volume ratio \SV{}. Surface relaxivity and
time-dependent diffusion are, in this sense, two readouts of one wall-collision
process on one substrate. A diffusion gradient resolves that substrate's
\emph{orientational} structure --- the fibre orientation distribution (FOD) --- and
its \emph{temporal} structure, $D(t)$; the gradient-free multi-echo spin echo reads
its scalar, orientation-integrated projection, the rate $\rho\,(S/V)$, fused into
what the experiment calls ``$T_2$''.

This shared origin touches both communities, and the usual assumption that diffusion is
``relaxivity-blind'' is only half true. A same-echo-time $b{=}0$ normalisation cancels the
\emph{common} surface factor $e^{-\mathrm{TE}\,\rho(S/V)}$, so the total signal and the fibre
orientation are indeed blind to it --- but intra- and extra-axonal water sit against
\emph{different} walls with different $S/V$, and that \emph{differential} $T_2$ does not
cancel. It reweights the compartments in the very $b{=}0$ that every diffusion microstructure
model normalises by, biasing the recovered \textbf{intra-axonal signal fraction}
$f_{\mathrm{intra}}$ (NODDI, spherical-mean, standard model). The same $S/V$ physics reads
through relaxometry as a bias in the \textbf{myelin water fraction (MWF)} --- the short-$T_2$
component of the multi-echo (CPMG / GRASE) decay, one of the most widely used myelin indices
\citep{Mackay1994, Laule2007, Whittall1997}, fit with a non-negative least-squares (NNLS)
$T_2$ spectrum and a fixed myelin-water window \citep{Whittall1997, Prasloski2012}. The
transverse rate the MWF is fit to is a \emph{material} bulk term $1/T_{2,\mathrm{bulk}}$ plus
the \emph{geometric} surface term $\rho\,(S/V)$ (Eq.~\eqref{eq:r2app}); both isotropic and
TE-linear, a non-identifiable combination no multi-echo measurement can separate
(Sec.~\ref{sec:theory:inseparability}). Surface relaxivity shortens the apparent
intra/extra-axonal $T_2$ toward the myelin window; for the thinnest axons it drags that $T_2$
\emph{below} the window, where the water is counted as myelin, so fine white matter of
\emph{identical myelin volume} reads myelin-\emph{richer}. Both effects are distinct from, and
additive to, the known microstructure dependences of these metrics --- $B_1$/flip angle
\citep{Prasloski2012, Lebel2010}, iron and susceptibility, and inter-compartment exchange
(itself an $S/V$-driven, axon-size-dependent MWF bias \citep{Harkins2012, Dula2010}); to our
knowledge neither has been reported as a surface-relaxivity term inseparable from the bulk
$T_2$.

We develop this on analytical footings and confirm it by simulation. We derive closed forms
for the interior (Brownstein--Tarr) and exterior (Novikov--Burcaw) surface rates over the
myelinated-cylinder FOD, and validate them with a wall-counting Monte Carlo that contains
\emph{no} analytical relaxivity model --- a parameter-free boundary-local-time estimator
consistent with the established simulator \textsc{MCMRSimulator} \citep{Cottaar2025}
(App.~\ref{sec:app:estimator}). We then carry the rate to the two microstructure metrics. For
the diffusion intra-axonal fraction, a closed-form packing law --- the interior/exterior $S/V$
ratio is $(1-\mathrm{VF})/(g\,\mathrm{VF})$ --- over-weights the intra-axonal signal (the
leading-order $f_{\mathrm{intra}}$ bias) by ${\approx}12\%$ on physiologically dense white
matter at clinical echo time, with a testable, packing-dependent TE drift. That compartment
fractions drift with TE when their $T_2$ differ is itself established \citep{Veraart2018}; our
contribution is to attribute that difference to a surface-relaxivity rate set by calibre and
packing, and the closed-form sign law that follows. For relaxometry, the same physics gives the
smaller MWF bias (fine WM reads myelin-richer, super-linear in the poorly-known $\rho$). Finally we ask whether the surface
term corrupts a longitudinal (lumen-preserving) demyelination measurement, and this is where the
g-ratio unifies the two metrics: the MWF bias is lumen-set (the interior wall), hence nearly
constant and \emph{cancels} in the change, whereas the $f_{\mathrm{intra}}$ bias --- set by the
interior-\emph{minus}-exterior differential --- \emph{drifts} strongly as the exterior wall
retreats with the myelin. One $S/V$ physics, longitudinally opposite readouts: the relaxometry
change is a cross-region confound only, but a diffusion $f_{\mathrm{intra}}$ change across a
demyelinating tract carries a large surface artefact.

Throughout we adopt the simplification the standard MWF analysis already makes ---
instantaneous, ideal refocusing pulses --- so that only bulk $T_2$ and surface
relaxivity are in play. In particular we exclude susceptibility, the \emph{other},
genuinely orientation-dependent fibre-angle dependence of MWF \citep{Birkl2021}; it and
finite-pulse coherence effects are orthogonal to the degeneracy shown here and treated
elsewhere.

\section{Theory}
\label{sec:theory}

\subsection{The myelin water fraction and surface relaxivity}
\label{sec:theory:mwf}

Myelin-water imaging reads the myelin water fraction (MWF) --- one of the most widely
used non-invasive indices of myelin content --- from the multi-echo spin-echo decay,
decomposing it with a non-negative $T_2$ spectrum into a short-$T_2$ pool (water trapped
between the myelin bilayers, $T_2\!\approx\!10$\,ms) and a long-$T_2$ intra/extra-axonal
(IE) pool ($T_2\!\approx\!50$--$80$\,ms), and reporting the short-$T_2$ signal fraction
\citep{Mackay1994, Whittall1997, Laule2007, Prasloski2012}.

The decay it fits, however, is not set by the molecular environment alone. A water
molecule diffusing among myelinated axons repeatedly strikes their walls, and each
collision shortens its transverse coherence --- \emph{surface relaxivity}
\citep{Brownstein1979} --- at a rate proportional to the pore surface-to-volume ratio
$S/V$. This geometric rate $\rho\,(S/V)$ adds to the bulk relaxation the fit is meant to
read, shortens the IE pool's apparent $T_2$ towards the myelin window, and is set by
axon calibre and packing --- microstructure the relaxometry experiment cannot resolve.
Two voxels of \emph{identical myelin volume} but different calibre or packing therefore
report different MWF. The rest of this section makes that precise: the surface term is
formally inseparable from the bulk $T_2$ (Sec.~\ref{sec:theory:inseparability}); it is
the relaxometric face of the same wall collisions diffusion MRI reads as time-dependent
diffusion (Sec.~\ref{sec:theory:dual}); and it has closed forms over the
myelinated-cylinder distribution (Sec.~\ref{sec:theory:closedforms}).

\subsection{The apparent transverse rate is non-identifiable}
\label{sec:theory:inseparability}

Two processes shorten the transverse magnetisation between the ideal refocusing pulses.
The first is intrinsic (\emph{bulk}) relaxation $1/T_{2,\mathrm{b}}$, set by the
molecular environment. The second is the surface loss introduced above ---
microscopically, paramagnetic surface sites and the field inhomogeneity of the interface:
in the fast-diffusion (motional-narrowing) limit, where a spin samples the whole pore
many times within an echo period, Brownstein and Tarr showed it is mono-exponential at a
rate proportional to the surface-to-volume ratio \citep{Brownstein1979},
\begin{equation}
  \frac{1}{T_{2,\mathrm{surf}}} \;=\; \rho\,\frac{S}{V},
  \label{eq:bt_rate}
\end{equation}
with $\rho$ the (transverse) surface relaxivity, a property of the wall.

The two processes add, so the rate the experiment actually decays at is
\begin{equation}
  R_2^{\mathrm{app}} \;=\; \frac{1}{T_{2,\mathrm{b}}} \;+\; \rho\,\frac{S}{V}.
  \label{eq:r2app}
\end{equation}
Crucially, the two terms of Eq.~\eqref{eq:r2app} are \emph{not separately observable in
a multi-echo measurement}, for two simultaneous reasons:

\begin{enumerate}
  \item \textbf{They share the same signal subspace.} Both terms are
  \emph{isotropic} --- surface relaxivity, like bulk relaxation, acts whenever the
  magnetisation is transverse, with no gradient or orientation dependence --- and both
  accrue \emph{at a constant rate in echo time}: in the motional-narrowing limit the
  surface attenuation is $\exp(-\rho\,(S/V)\,t)$, the same single-exponential form as
  $\exp(-t/T_{2,\mathrm{b}})$. For a single pore size they are exactly degenerate; a
  pore-size \emph{distribution} makes the surface term mildly multi-exponential, but its
  leading effect is a mean shift $\rho\,\langle S/V\rangle$ that is itself degenerate
  with the bulk rate, and the residual curvature does not lift the degeneracy in a
  standard fit.

  \item \textbf{The geometry term is unobserved.} The surface term depends on $S/V$, a
  property of the sub-voxel geometry that a relaxometry experiment does not resolve ---
  many substrates of different calibre or packing give near-identical decays --- so
  $\rho\,(S/V)$ cannot be predicted and subtracted, and $\rho$ itself is not
  independently established for myelin. It is an \emph{unobserved} addition to what the
  fit reports as $1/T_2$.
\end{enumerate}

In short: what the fit calls $T_2$ is already a mixture of the material
$1/T_{2,\mathrm{b}}$ and a geometric $\rho\,(S/V)$ that no multi-echo measurement can
disentangle --- and this holds even with no diffusion-encoding gradient, since the
spins strike walls regardless.

\subsection{One wall-collision process, two observables}
\label{sec:theory:dual}

The surface term in Eq.~\eqref{eq:r2app} is not an exotic addition; it is the
quantitative-MRI face of a process the diffusion-MRI community already
characterises in detail. Novikov, Fieremans, Burcaw and colleagues
\citep{Novikov2014, Burcaw2015, NovikovKiselev2010} showed that in randomly packed media the
\emph{extra-axonal} diffusion coefficient is time-dependent: as the diffusion
time grows, walkers encounter more of the disordered packing, and the
ensemble-averaged $D(t)$ relaxes towards its tortuosity limit $D_\infty$ (the
long-time plateau diffusivity) at a rate set by the two-point density correlation
function $\Gamma(\mathbf r)$ --- how the cylinder positions correlate across the pack ---
\begin{equation}
  D(t) \;\approx\; D_\infty + \frac{\Gamma_0}{2\pi}\,\frac{\ln(4 D_\infty t/e\zeta^2)}{t}
  \qquad\text{(2D random cylinder packing)},
  \label{eq:dt}
\end{equation}
with $\Gamma_0$ and $\zeta$ the disorder strength and correlation length. The
diffusion experiment reads this off as a $b$-value-encoded, time-dependent
attenuation.

The same wall collisions that produce the time dependence of $D(t)$ are, from the
qMRI side, surface-relaxivity events. \emph{Every} encounter of a spin with a wall
both (i) interrupts its free displacement --- the event whose statistics, summed
over the packing, give $D(t)$ --- and (ii) exposes it to the wall's relaxation
sink, accumulating surface local time. It is the \emph{same walls}, counted two
ways. How tightly the two rates share a functional object depends on the regime. At
\emph{short} times both are governed by the small-wavevector limit of the same packing
correlation function $\Gamma(\mathbf r)$: the Mitra $t^{-1/2}$ surface term of the
relaxation is the literal twin of the $\sqrt{t}$ correction to $D(t)$
\citep{Mitra1992, NovikovKiselev2010}. At the \emph{long} times a myelin-water train runs,
however, $D(t)$ still carries $\Gamma$'s shape (its approach to the tortuosity plateau,
Eq.~\eqref{eq:dt}) whereas the surface rate has saturated to the motional-narrowing
plateau $\rho\,\langle S/V\rangle$ --- set by the \emph{mean} exterior surface density
(the normalisation of $\Gamma$), not its shape. So the honest statement is not ``one
correlation function, two observables'' at all times, but: two readouts of one
wall-collision process on one substrate, sharing a common short-time origin and a common
mean geometry. The diffusion measurement plays a gradient and so resolves that
substrate's \emph{orientational} structure (the fibre orientation distribution) and its
\emph{temporal} structure $D(t)$; the relaxometry measurement plays no gradient and reads
only the scalar, orientation-integrated projection, the rate $\rho\,(S/V)$.

This is literal in the spherical-convolution picture of white matter. The voxel is
one fibre orientation distribution (FOD) $\mathcal F(\hat{\mathbf n})$ convolved with
one per-fibre response kernel $\mathcal K$ \citep{tournier2007},
\begin{equation}
  S(\hat{\mathbf g},b,\mathrm{TE})=\int_{S^2}\mathcal F(\hat{\mathbf n})\,
  \mathcal K(\hat{\mathbf g}\!\cdot\!\hat{\mathbf n};b,\mathrm{TE})\,\mathrm d\hat{\mathbf n},
  \qquad
  \mathcal K=\!\!\sum_{c\in\{\mathrm i,\mathrm e,\mathrm m\}}\!\! f_c\,
  e^{-\mathrm{TE}\,R_{2,c}}\,E_c(\hat{\mathbf g}\!\cdot\!\hat{\mathbf n},b),
  \label{eq:fod_conv}
\end{equation}
each compartment carrying its transverse rate $R_{2,c}$ ---
$R_{2,\mathrm i}=1/T_{2,\mathrm b}+\rho\,\svin$ (interior wall),
$R_{2,\mathrm e}=1/T_{2,\mathrm b}+\rho\,S_{\mathrm{ext}}/V_{\mathrm{ext}}$ (exterior;
Sec.~\ref{sec:theory:closedforms}) --- and its diffusion attenuation $E_c$. The
surface-relaxivity rate is gradient-free, so it sits entirely in the $\ell=0$ monopole of
this kernel, $\kappa_0(\mathrm{TE})=\sum_c f_c\,e^{-\mathrm{TE}R_{2,c}}$: a multi-echo
experiment carries the surface term in full (fused into the $R_{2,c}$) but is blind to
the FOD's shape, which is why relaxometry reports a scalar $\rho\,(S/V)$ with no
fibre-angle dependence (absent susceptibility).

Standard diffusion processing, conversely, normalises that factor away. A diffusion
experiment divides by its own $b{=}0$ image at the \emph{same} echo time, so the common
relaxation factor $e^{-\mathrm{TE}(1/T_{2,\mathrm b}+\rho S/V)}$ cancels and the FOD it
estimates is, to leading order, \emph{relaxivity-blind} --- exact for a single
compartment, with a multi-compartment voxel leaving only a second-order TE-reweighting of
the compartment fractions,
\begin{equation}
  E(\hat{\mathbf g},b)=\sum_c w_c(\mathrm{TE})\,E_c(\hat{\mathbf g},b),
  \qquad
  w_c(\mathrm{TE})=\frac{f_c\,e^{-\mathrm{TE}\,R_c}}{\sum_{c'} f_{c'}\,e^{-\mathrm{TE}\,R_{c'}}},
  \label{eq:te_weights}
\end{equation}
the same effect that makes the apparent intra-axonal fraction TE-dependent
\citep{Veraart2018}. The surface term is thus the one substrate property that relaxometry
cannot separate from the bulk $T_2$ (Sec.~\ref{sec:theory:inseparability}) and that
diffusion divides out --- carried by both signals, resolved by neither, yet set by the
same wall geometry. Whether a tailored diffusion acquisition could nonetheless recover
that geometry is a separate question we return to, and qualify, in the Discussion.

\subsection{Closed forms over the myelinated-cylinder distribution}
\label{sec:theory:closedforms}

A myelinated axon is two concentric walls: the spins of the intra-axonal
compartment diffuse in the lumen and bounce on the \emph{inner} wall at the inner
(axonal) diameter $d$, while the extra-axonal spins bounce on the \emph{outer}
myelin surface at diameter $d/g$, with $g$ the g-ratio. White-matter inner
diameters follow a Gamma distribution $P(d)\propto d^{\alpha-1}e^{-d/\beta}$ with
shape $\alpha\!\approx\!2$ \citep{Aboitiz1992, Assaf2008}; the outer-diameter
distribution is the same scaled by $1/g$. The two walls therefore set two
contributions to the surface-relaxivity attenuation, one keyed to $d$ and one to
$d/g$.

\paragraph{Interior wall (intra-axonal water).} For a cylinder of inner diameter
$d$, the surface-to-volume ratio is $S/V=4/d$. Spins are populated by
cross-sectional area ($\propto d^2$), so the compartment attenuation is the
\emph{area-weighted} (``spin-weighted'') average of the Brownstein--Tarr factor
over $P(d)$ --- the same population weighting the AxCaliber diffusion signal uses
\citep{Assaf2008}. With $S/V=4/d$ the relevant moment is
\begin{equation}
  \Big\langle \tfrac{4}{d}\Big\rangle_{V}
  \;=\; \frac{\int 4/d\;d^2 P(d)\,\mathrm dd}{\int d^2 P(d)\,\mathrm dd}
  \;=\; \frac{4}{\beta(\alpha+1)},
  \label{eq:area_moment}
\end{equation}
so the area-weighted attenuation is a modified-Bessel form in $t$ with effective
rate $\rho\,\svin$.

\paragraph{Exterior wall (extra-axonal water).} Extra-axonal spins bounce on the
outer (myelin) surfaces, at diameter $d/g$, of the surrounding cylinders --- so the
exterior surface density $S_{\mathrm{ext}}/V_{\mathrm{ext}}$ carries the g-ratio.
Its surface-relaxivity rate is governed by the \emph{same} disordered packing that sets
the time-dependent diffusion of Eq.~\eqref{eq:dt} --- sharing its short-time
$\Gamma$-controlled behaviour and, as we now show, saturating at long times to a plateau
fixed by the mean exterior surface density.

At \emph{short} diffusion times the exterior rate follows the Mitra $t^{-1/2}$
surface-to-volume law \citep{Mitra1992, Mitra1993} from the same small-wavevector
correlations $\Gamma(k)\sim k^p$ ($p=0$ for randomly packed cylinders) that set $D(t)$
\citep{NovikovKiselev2010} --- the short-time relaxometric twin of the Mitra $D(t)$
expansion. Multi-echo myelin-water trains, however, run to \emph{long} echo times
($t\gg\zeta^2/D_\infty$) in the \emph{fast-diffusion} (motional-narrowing) regime
\citep{Brownstein1979}, where $\rho\,\ell/D_\infty\ll1$; both hold comfortably for
myelinated white matter (with $\zeta\!\sim\!1\,\mu$m and $D_\infty\!\sim\!1\,\mu$m$^2$/ms
the crossover $\zeta^2/D_\infty\!\sim\!1$\,ms sits far below the $10$--$320$\,ms echo
times, and $\rho\,\ell/D_\infty\!\approx\!10^{-3}$). A walker therefore samples the full
exterior geometry many times per echo, the rate saturates to a constant plateau,
\begin{equation}
  \frac{1}{T_{2,\mathrm{surf}}^{\mathrm{ext}}}
  \;=\; \rho^{\mathrm{ext}}\,\frac{S_{\mathrm{ext}}}{V_{\mathrm{ext}}},
  \label{eq:b_ext_long}
\end{equation}
identical in form to the interior term and to Eq.~\eqref{eq:bt_rate}. Two distinct
statistical objects are at play here, and it is worth separating them cleanly. The long-time
plateau rate is a \emph{one-point} quantity --- the \emph{mean} exterior surface density
$S_{\mathrm{ext}}/V_{\mathrm{ext}}$ (number density $\times$ per-cylinder outer perimeter over
extra-axonal area). It equals $\rho\,\langle S/V\rangle$ because, in the fast-diffusion limit,
the lowest Robin eigenvalue of the connected extra-axonal domain reduces to $\rho$ times its
total surface over total volume whenever the corresponding eigenmode is uniform across the pore
network --- i.e.\ under motional narrowing over the whole extra-axonal space. The \emph{approach}
to that plateau, by contrast, is governed by the \emph{two-point} density correlation
$\Gamma(\mathbf r)$ (Eq.~\eqref{eq:dt}) --- the same object that sets the time-dependence of
$D(t)$, and whose small-wavevector limit gives the short-time Mitra twin. So the two observables
share the two-point $\Gamma$ in their \emph{time-dependence} and share only the one-point mean
geometry at the \emph{plateau} the multi-echo train integrates over. The mean-field plateau is
moreover exact only when the extra-axonal water motionally averages over the \emph{local}
distribution of $S/V$ within an echo; where tight near-contacts kinetically isolate high-$S/V$
pockets the uniform-eigenmode assumption fails and the arithmetic mean is a lower bound (by
Jensen, the convex Robin response to a distribution of local $S/V$ exceeds the response to the
mean), a limit we return to in Sec.~\ref{sec:discussion}. Both walls therefore contribute an
isotropic, TE-linear Brownstein--Tarr rate to Eq.~\eqref{eq:r2app}, tied through $d$ and $g$ to
the substrate geometry the relaxometry experiment does not resolve.

For the ratio of the two walls the calibre distribution drops out entirely. Writing the
exterior density as total outer-wall perimeter over extra-axonal area,
$S_{\mathrm{ext}}/V_{\mathrm{ext}}=\sum 2\pi r_{\mathrm{out}}/(L^2-\sum\pi r_{\mathrm{out}}^2)$,
and using $r_{\mathrm{in}}=g\,r_{\mathrm{out}}$ with the fibre (outer) volume fraction
$\mathrm{VF}=\sum\pi r_{\mathrm{out}}^2/L^2$, the interior-to-exterior ratio reduces to a
purely geometric factor,
\begin{equation}
  \frac{\svin}{S_{\mathrm{ext}}/V_{\mathrm{ext}}}
  \;=\; \frac{1-\mathrm{VF}}{g\,\mathrm{VF}},
  \label{eq:svratio_theory}
\end{equation}
\emph{independent of the calibre distribution}, which sets only the common scale. The two
rates cross at $\mathrm{VF}^{\ast}=1/(1+g)$: below it the interior wall dominates, above it
the extra-axonal water --- squeezed into thin, high-$S/V$ crevices between tightly packed
fibres --- relaxes faster. This single ratio governs the intra-axonal signal-fraction bias
of Sec.~\ref{sec:results:fintra}.

\section{Methods}
\label{sec:methods}

A wall-counting Monte Carlo shows the multi-echo rate absorbs $\rho\,(S/V)$ with no
gradient (Sec.~\ref{sec:methods:mc}); that validated rate is then carried, over the full
calibre distribution, into two microstructure readouts on a fixed-myelin substrate --- the
diffusion intra-axonal signal fraction (Sec.~\ref{sec:methods:fintra}) and the standard
myelin-water analysis (Sec.~\ref{sec:methods:mwf}). All constants are in
Table~\ref{tab:constants}.

\subsection{Canonical white-matter substrate}
\label{sec:methods:substrate}

\begin{figure*}[t]
  \centering
  \includegraphics[width=0.96\linewidth]{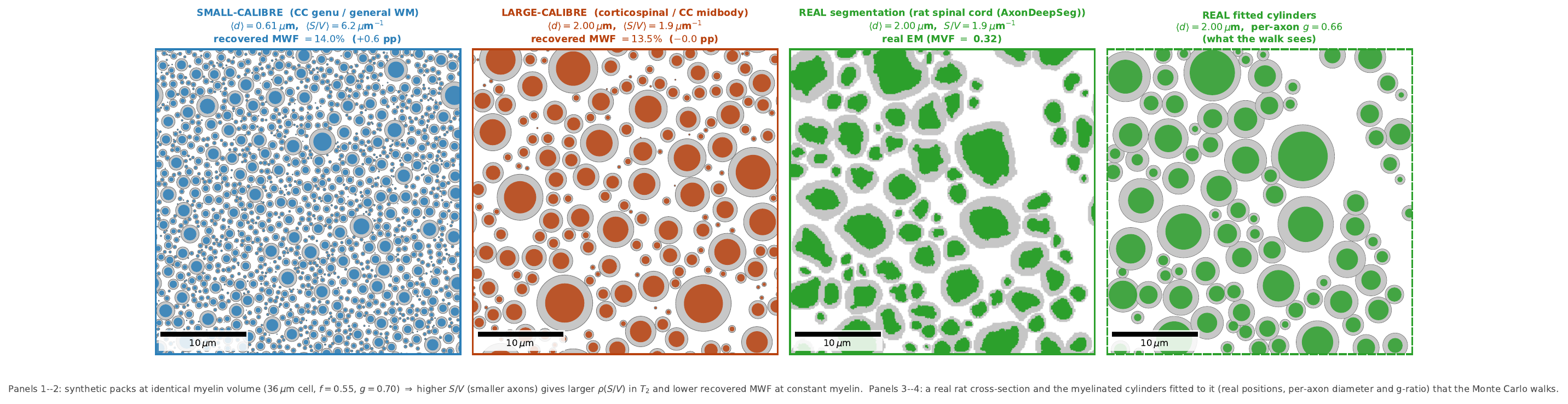}
  \caption{\textbf{White-matter substrates: synthetic packs at fixed myelin volume,
  and a real cross-section beside the cylinders fitted to it.}
  \emph{Panels 1--2:} small- and large-calibre random packs of Gamma-distributed
  myelinated cylinders at the same scale, fibre fraction $f=0.55$, g-ratio $g=0.70$, and
  \emph{the same myelin volume}; small axons pack more wall per unit volume (higher
  $S/V$), so the absorbed $\rho\,(S/V)$ is larger and the recovered MWF (annotated, from
  the Fig.~\ref{fig:mwf} forward model) is lower at identical myelin content.
  \emph{Panels 3--4:} a real rat spinal-cord cross-section (AxonDeepSeg, $0.13\,\mu$m/px;
  \citealp{Zaimi2018}) and the myelinated cylinders fitted to it
  (Sec.~\ref{sec:methods:real}; dashed border) --- one of the $10$ cross-sections whose
  $S/V$ range is the rug in Fig.~\ref{fig:mwf}B.}
  \label{fig:substrate}
  {\footnotesize\textit{Reproducible from} \texttt{figures/fig\_substrate\_cross\_section.py}.}
\end{figure*}

We act on three substrates (Fig.~\ref{fig:substrate}): two synthetic packs that
bracket the physiological calibre range and one real rat cross-section. Each
synthetic substrate is a random pack of myelinated cylinders with Gamma-distributed
inner diameters (shape $\alpha=2$), g-ratio $g=0.70$, and fibre (axon$+$myelin)
volume fraction $f=0.55$, giving a myelin volume fraction
$\mathrm{MVF}=f(1-g^2)=0.28$ and, with a myelin water content $w_{\mathrm m}=0.40$
\citep{West2018}, a true myelin water signal fraction of $13.5\%$
\citep{Aboitiz1992, Assaf2008, Fieremans2011}. \emph{The myelin volume is held
fixed throughout.} Axon calibre is swept by changing the Gamma scale (mean inner
diameter $0.5$--$3\,\mu$m) at fixed $g$ and $f$, so the surface-to-volume ratio
varies by a factor of six while the myelin volume fraction is identically
constant. The relevant interior moment is the \emph{area/spin-weighted}
$\svin=4/[\beta(\alpha+1)]$ (Sec.~\ref{sec:theory:closedforms}),
not the bare $4/d$. The reference relaxivity $\rho=1.16\,\mu$m/s is a
histology-calibrated model coefficient \citep{Barakovic2023}, not an independently
measured axolemma relaxivity, so we sweep $\rho$ over $0$--$2.5\,\mu$m/s and report the
result as a bound (Sec.~\ref{sec:results:mwf}).

\begin{table}[t]
\centering
\small
\caption{Constants used. White-matter values are the canonical substrate of the
companion modelling work; sources as cited. Orientation dispersion is listed for
completeness: being orientation-blind it does \emph{not} affect surface relaxivity
or the MWF bias.}
\label{tab:constants}
\begin{tabular}{@{}lll@{}}
\toprule
Symbol & Quantity & Value \\
\midrule
$\alpha$ & Gamma diameter shape & $2$ \\
$d$ & mean inner diameter (swept) & $0.5$--$3.0\,\mu$m \\
$g$ & g-ratio \citep{Stikov2015} & $0.70$ \\
$f$ & fibre volume fraction & $0.55$ \\
$\mathrm{MVF}$ & myelin volume fraction ($=f(1-g^2)$, fixed) & $0.28$ \\
$w_{\mathrm m}$ & myelin water content \citep{West2018} & $0.40$ \\
$\rho$ & transverse surface relaxivity \citep{Barakovic2023} & $1.16\,\mu$m/s \\
$T_{2,\mathrm{b}}^{\mathrm{IE}}$ & intrinsic (surface-free) IE bulk $T_2$ & $75$\,ms \\
$T_2^{\mathrm{MW}}$ & myelin water $T_2$ & $12$\,ms \\
$D_0$ & intrinsic diffusivity & $1.7$--$2.0\,\mu$m$^2$/ms \\
$\mathrm{ODI}$ & orientation dispersion index, typical WM \citep{Zhang2012,Tariq2016}
  & $0.1$--$0.2$ \\
\multicolumn{2}{@{}l}{CPMG: $32$ echoes, $\mathrm{TE}=10$\,ms, ideal $180^\circ$} & \\
\bottomrule
\end{tabular}
\end{table}

\subsection{Monte Carlo: surface relaxivity by wall counting}
\label{sec:methods:mc}

The simulator (dmipy-sim) propagates $4\times10^{4}$ random walkers through the
explicit packed geometry under Brownian dynamics, with specular reflection at the
cylinder walls. Surface relaxation is applied inline during propagation: at each
wall encounter the walker's transverse magnetisation is multiplied by
$\exp(-2\rho\,d_\perp/D)$, where $d_\perp$ is the perpendicular overshoot of the
reflected step, and these per-encounter factors accumulate into a walker weight
$\exp(-\rho\,\mathcal L_{\partial\Omega}/D)$ with $\mathcal L_{\partial\Omega}$ the
total wall contact --- a realised-overshoot estimator of the boundary local time that
realises the Brownstein--Tarr Robin condition without any analytical relaxivity model in
the propagation (App.~\ref{sec:app:estimator} derives the local-time construction,
verifies it against the closed form, and confirms it is consistent with the established
estimator of \citealp{Cottaar2025}). The local time is resolved by sub-stepping the walk
to the extra-axonal pore scale ($\mathrm{step}\!\lesssim\!(S_{\mathrm{ext}}/V)^{-1}/8$):
the confined intra-axonal walkers fully sample the inner wall and are accurate at any step,
but the fast extra-axonal walkers under-count grazing outer-wall contact at a coarse step,
so both compartments converge onto the closed form only when the extra pore is resolved. The
Brownstein--Tarr behaviour, if it emerges, emerges from
counting wall encounters. The myelin is
made impermeable ($\kappa=0$) so that the intra- and extra-axonal compartments are
not contaminated by exchange into the short-$T_2$ myelin pool, isolating surface
relaxivity as the only wall effect. The pure surface factor of each compartment is
the mean surviving weight
$\langle\exp(-\rho\,\mathcal L_{\partial\Omega}/D)\rangle$ over that compartment's
walkers (no diffusion gradient, no bulk-$T_2$ term), and compared against its closed
form --- the area/spin-weighted Brownstein--Tarr interior term and the
Novikov--Burcaw exterior term --- over the realised pack, versus echo time at the
canonical substrate (Fig.~\ref{fig:mwf}A). As a unit check, the same wall-counting
on a single cylinder of radius $R$ reproduces the Brownstein--Tarr rate $\rho\,(2/R)$
to $<1\%$ across radii at the production sub-step.

\subsection{Intra-axonal signal-fraction bias}
\label{sec:methods:fintra}

The diffusion intra-axonal fraction $f_{\mathrm{intra}}$ is estimated from signal
normalised by a $b{=}0$ at the same echo time, under the assumption of a shared
compartment $T_2$; surface relaxivity breaks that assumption by giving the interior and
exterior walls different $S/V$ (Sec.~\ref{sec:theory:closedforms}), hence different
apparent $T_2$. We compute the resulting bias two ways, which agree to
${\lesssim}2$\,pp. The \emph{exact} calculation carries the full calibre distribution: at
diffusion echo times the myelin water has decayed, so the two surviving compartments are
the intra-axonal lumen (volume $\propto\mathrm{VF}\,g^2$, relaxing over the distributed
interior rate $\rho\,(4/d)$ weighted by water volume $\propto d^2 P(d)$) and the
extra-axonal water (volume $\propto 1-\mathrm{VF}$, at the exterior rate
$\rho\,S_{\mathrm{ext}}/V_{\mathrm{ext}}$); the apparent fraction is the TE-weighted ratio
of these attenuated amplitudes. The \emph{closed form} replaces the distributed intra pool
by its mean rate, giving the single-exponential odds rescaling of Eq.~\eqref{eq:fbias}
governed by the calibre-independent ratio Eq.~\eqref{eq:svratio_theory}. Both are evaluated
over fibre volume fraction ($0.30$--$0.85$), calibre, and echo time
(Fig.~\ref{fig:fintra}); no estimator or Monte-Carlo replay is involved, the bias being an
analytic consequence of the two validated wall rates.

\subsection{Spherical-mean fit and the injected-prior correction}
\label{sec:methods:smt}

To confirm the signal-weight bias survives an actual estimator (Fig.~\ref{fig:smt}), we
forward-simulate a dense white-matter multi-shell diffusion signal ($\mathrm{VF}=0.68$,
$3$ shells $b=1,2,3\,$ms/$\mu$m$^2$, $90$ directions each plus $8$ $b{=}0$, PGSE
$\delta/\Delta=12/40\,$ms) from the analytical white-matter model with the surface factors on,
at echo times $50$--$90\,$ms, and fit each with the standard spherical-mean technique (SMT: a
stick intra-axonal $+$ zeppelin extra-axonal spherical-mean model \citep{Kaden2016}) in
\texttt{dmipy-fit}. The compartment diffusivities are fixed to their known values (leaving the
fraction the only free parameter), because a free-diffusivity fit is degenerate on this heavily
$T_2$-reweighted signal. To isolate the \emph{surface} bias from the SMT model-mismatch
(a two-compartment stick$+$zeppelin fit to a three-compartment substrate) we fit the
\emph{same} model to a surface-off ($\rho=0$) forward at each TE; the surface-off recovered
fraction is TE-independent (spread ${<}0.01$ across $50$--$90\,$ms), confirming the TE drift of
the surface-on fit is the surface term and not a fitting artefact. The injected-prior
correction takes the region's histology calibre prior (fixed $\alpha$, outer scale, $g$,
$f_{\mathrm{axon}}$, $\rho$), from which $\rho\,[(S/V)_{\mathrm{int}}-(S/V)_{\mathrm{ext}}]$ is
known, and inverts the odds rescaling of Eq.~\eqref{eq:fbias} at each single TE.

\subsection{Instantaneous-pulse CPMG and the standard MWF estimate}
\label{sec:methods:mwf}

We adopt the standard multi-echo protocol ($32$ echoes, $\mathrm{TE}=10$\,ms) with
ideal, instantaneous $180^\circ$ refocusing --- the conventional assumption, under which
the only transverse-decay mechanisms are bulk relaxation and surface relaxivity. The MWF
is estimated with the standard chi-square-regularised NNLS $T_2$ spectrum (zero-order
Tikhonov weight set per signal within a factor $1.02$ of the unregularised misfit, the
Whittall--MacKay\,/\,Prasloski\,/\,DECAES choice; myelin signal integrated below a
$25$\,ms cutoff) \citep{Whittall1997, Prasloski2012, Doucette2020} --- the same basic NNLS
optimisation the standard pipelines use. The intrinsic IE bulk $T_2$
is fixed at $T_{2,\mathrm{b}}^{\mathrm{IE}}=75$\,ms (a single material constant), which at
the canonical $S/V$ reproduces the literature apparent IE $T_2$ of $\sim$$50$--$70$\,ms
once the surface term is added.

The decay fed to the estimator is a fixed-myelin forward model carrying the
Monte-Carlo--validated rate (Sec.~\ref{sec:methods:mc}) as its only surface input; using
a forward model rather than raw simulated signal isolates the surface confound from
protocol-specific spectral artefacts and, crucially, \emph{retains the full calibre
distribution} that the simulator's step-size truncation removes from the raw walk. There
is a myelin-water pool ($T_2^{\mathrm{MW}}=12$\,ms), an extra-axonal pool at
$1/T_{2,\mathrm{b}}^{\mathrm{IE}}+\rho\,(S_{\mathrm{ext}}/V)$, and --- the key ingredient
--- a \emph{calibre-distributed intra-axonal pool}: each inner diameter $d$ in the Gamma
population relaxes at $1/T_{2,\mathrm{b}}^{\mathrm{IE}}+\rho\,(4/d)$, water-volume-weighted,
so the intra signal is a spectrum rather than a single exponential (a lumped mean rate
misses the thin-calibre tail entirely, and reverses the sign of the recovered bias). The
amplitudes are fixed by the fixed myelin volume: with fibre fraction $f=0.55$, $g=0.70$ and
myelin water content $w_{\mathrm m}=0.40$, the water-weighted volumes are myelin
$f(1-g^2)w_{\mathrm m}=0.112$, intra (lumen) $f g^2=0.270$, and extra $1-f=0.45$, normalising
to the signal fractions $f_{\mathrm m}=0.135$, $f_{\mathrm{ia}}=0.324$,
$f_{\mathrm{ea}}=0.541$. The bulk $T_2$ is a fixed material constant, and only the geometric
$S/V$ (through calibre) and the unknown $\rho$ vary. We assign intra and extra the \emph{same}
intrinsic bulk $T_2$, so their apparent difference is the surface term alone; a genuine
intrinsic intra/extra $T_2$ difference would add to it, and our bias is the surface
contribution on top of any such offset. The
robust readout needs no estimator: the short-$T_2$ signal fraction --- myelin water plus
the intra water with $T_2$ below the $25$\,ms cutoff, i.e.\ from axons thinner than
$d^{\ast}=4\rho/(1/T_2^{\mathrm{cut}}-1/T_{2,\mathrm{b}}^{\mathrm{IE}})$ --- is a grid-free
property of the decay, which we report alongside the standard NNLS (which reproduces its
sign and amplifies it). Robustness is checked under the chi-square-adaptive weight, fixed
zero-order weights ($0.01$, $0.02$), and Rician noise (single-voxel $\mathrm{SNR}=150$).

\subsection{End-to-end check: MWF from the full Monte-Carlo signal}
\label{sec:methods:engine}

As an end-to-end check, free of any forward model, we simulate the full multi-echo CPMG
decay with the forward Monte Carlo (instantaneous $180^\circ$; a per-echo
gradient-free walk carrying per-compartment bulk $T_2$ and first-principles wall relaxivity,
no analytical decay) and run the same NNLS estimator on it. The intrinsic bulk $T_2$ is a
fixed material value ($75$\,ms) at the actual walls, so turning surface relaxivity on adds
$\rho\,(S/V)$ and produces the calibre-dependent apparent decay. Over the long CPMG train the
surface local time is sub-stepped at the extra-axonal pore/2 (Sec.~\ref{sec:methods:mc}), a
coarser resolution than the Panel-A validation but ample for the decay. We walk four calibres
(surface on and off on each), add Rician noise at $\mathrm{SNR}=200$ ($150$ realisations), and
NNLS-fit each. This end-to-end walk reproduces the decay and the IE-pool spectral shift, but
its \emph{integrated} short-$T_2$ fraction barely moves: we deliberately do not walk the
sub-micron crossing calibres here (below ${\sim}0.4\,\mu$m an impermeable-wall walk needs
prohibitively small steps and compute), so the quantitative MWF bias is carried by the
full-distribution forward model above (Sec.~\ref{sec:methods:mwf}).

\subsection{Real white-matter cross-sections}
\label{sec:methods:real}

To anchor the substrate in real tissue and to test that the operating point was not
cherry-picked, we use every manually-segmented cross-section in the
AxonDeepSeg SEM dataset (rat spinal cord, $8$ animals, $10$ cross-sections,
$0.07$--$0.18\,\mu$m/px; \citealp{Zaimi2018}). Each cross-section is represented as
a pack of myelinated cylinders fitted to its segmentation: every segmented axon
becomes a cylinder at its true centroid, with inner radius the area-equivalent
radius of the lumen and outer radius the area-equivalent radius of the lumen plus
its watershed-assigned myelin ring, giving a \emph{per-axon} diameter and g-ratio.
This preserves the geometry the physics depends on --- the axon-size distribution
and packing, hence $S/V$ --- while routing the walk through the validated exact-circle
reflection (the irregular-segmentation walk does not confine the extra-axonal
walkers; the fitted-cylinder representation does). The fitted reconstruction is
shown beside the raw segmentation in Fig.~\ref{fig:substrate} for transparency. The
interior area-weighted moment $\svin$ across the $10$ cross-sections
spans $0.9$--$2.3\,\mu$m$^{-1}$ (the rug in Fig.~\ref{fig:mwf}B), confirming that
real white matter populates the swept $S/V$ range.

\subsection{Demyelination trajectory}
\label{sec:methods:demyel}

To ask how the surface-relaxivity bias behaves as a tract demyelinates, we trace a
\emph{within-subject} series rather than compare independently sampled substrates. For
each of $N_{\mathrm{subj}}=24$ ``subjects'' we draw a Gamma axon population
($\alpha=2$, canonical calibre, $N=300$ cylinders) and pack the outer (myelin) cylinders
\emph{once}, at the baseline g-ratio $g_0=0.70$ --- the thickest myelin, hence the
largest outer walls; packing the largest-ever cylinders first guarantees every later,
thinner state is overlap-free. We then \emph{freeze the centres and the inner (axon)
radii} and demyelinate by raising $g$ toward $1$ at fixed axon, so the outer wall shrinks
($r_{\mathrm{out}}=r_{\mathrm{in}}/g$) while the axon count and lumen are preserved. Each
subject is thus its own longitudinal series on one fixed axon distribution, not a fresh
random substrate per stage. At every stage we read the geometry directly off the realised
pack --- the myelin volume fraction
$\mathrm{MVF}=\sum\pi(r_{\mathrm{out}}^2-r_{\mathrm{in}}^2)/L^2$ (hence the true MWF), the
interior $\svin$ (unchanged, the axon lumen being preserved), and the
exterior $S_{\mathrm{ext}}/V_{\mathrm{ext}}=\sum 2\pi r_{\mathrm{out}}/(L^2-\sum\pi r_{\mathrm{out}}^2)$
(which falls as the outer walls retreat) --- and feed the \emph{frozen inner-radius
distribution} to the same calibre-distributed forward model as Sec.~\ref{sec:methods:mwf},
read as the grid-free short-$T_2$ fraction, at $\rho=0$ (true MWF) and at the cited $\rho$
(apparent MWF), averaging over subjects. Because the crossing that sets the bias depends on
the inner radii --- which this trajectory holds fixed --- the offset is nearly constant and
cancels in the within-subject change; the exterior $S/V$ that halves plays no role in it. No new wall
rate is introduced: the surface rate is the one the Monte Carlo validated in
Sec.~\ref{sec:methods:mc}, applied to the demyelinating geometry. The $24$ packs differ
only in random cylinder placement, so the spread across them measures packing-realisation
variance (negligible here) rather than biological variability; the trajectory is, in
effect, deterministic.

\section{Results}
\label{sec:results}

\subsection{Surface relaxivity enters the signal and the apparent $T_2$}
\label{sec:results:signal}

\begin{figure}[t]
  \centering
  \includegraphics[width=\linewidth]{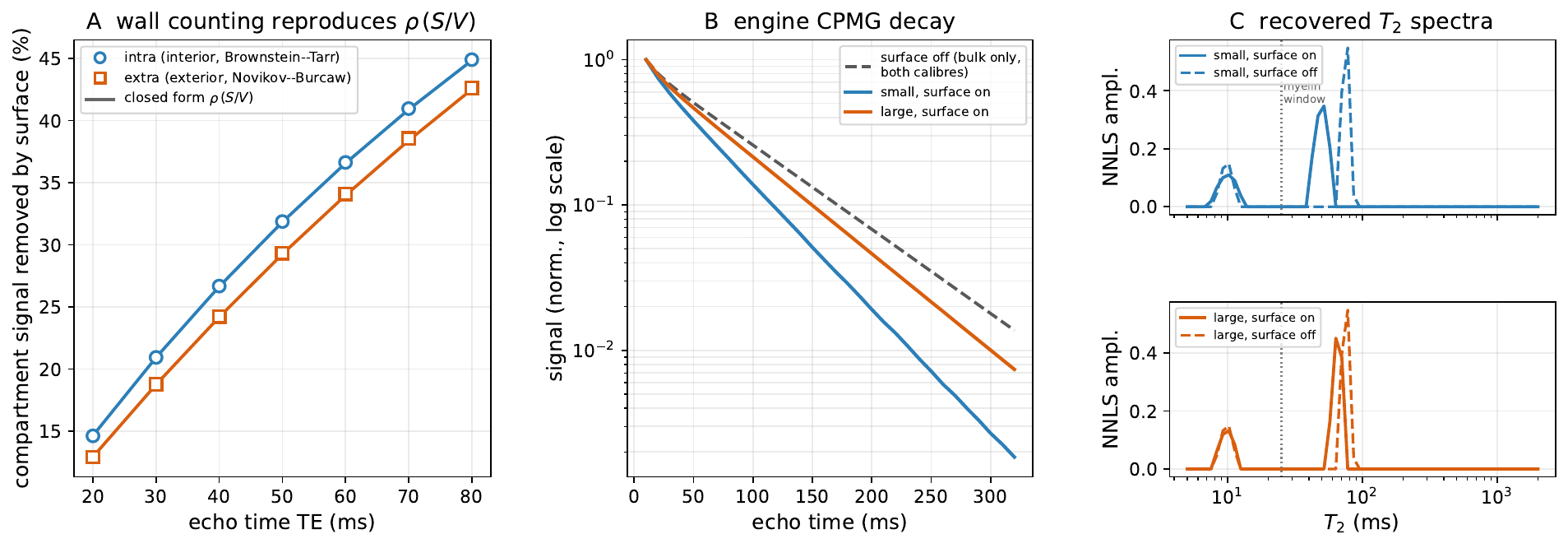}
  \caption{\textbf{Surface relaxivity enters the multi-echo signal and the apparent
  $T_2$.}
  \textbf{(A)} Forward Monte Carlo surface attenuation at $b=0$ (wall counting, no
  analytical relaxivity model; markers) versus the closed-form interior (Brownstein--Tarr,
  circles) and exterior (Novikov--Burcaw, squares) factors (solid lines) --- both walls
  reproduced to ${\lesssim}0.2$\,pp (the engine sub-steps the boundary local time to the
  extra-axonal pore scale). So wall-counting reproduces $\rho\,(S/V)$, and the multi-echo
  $T_2$ carries it with no gradient.
  \textbf{(B)} Engine-simulated gradient-free CPMG decay (log scale), surface on (solid)
  for small/large calibre vs the calibre-independent surface-off reference (dashed grey):
  the surface term speeds the decay, more at high $S/V$.
  \textbf{(C)} Standard-NNLS $T_2$ spectra, surface on vs off: surface relaxivity shifts
  the long-$T_2$ IE pool toward the $25$\,ms myelin window, more at high
  $S/V$.}
  \label{fig:signal}
  {\footnotesize\textit{Reproducible from} \texttt{figures/fig\_cpmg\_mwf\_engine.py} (panel~A reads \texttt{fig\_mwf\_sv\_bias.py}).}
\end{figure}

Figure~\ref{fig:signal}A shows the surface attenuation of the intra- and
extra-axonal water at $b=0$ --- with no diffusion-encoding gradient --- as a
function of echo time, on the canonical substrate. The simulator contains no
analytical relaxivity model: it propagates walkers, reflects them specularly at the
walls, and accumulates the wall-contact local time, so the surface attenuation
$\langle\exp(-\rho\,\mathcal L_{\partial\Omega}/D)\rangle$ read off the recorded
contact is the simulator's first-principles wall dephasing. It lands on the
closed-form surface factor for both walls --- the interior Brownstein--Tarr term
over the realised inner-diameter distribution and the exterior Novikov--Burcaw
term (Fig.~\ref{fig:signal}A), removing $36.4\%$ (Monte Carlo) versus $36.6\%$ (closed form) of
the intra-axonal amplitude and $34.0\%$ versus $34.0\%$ of the extra-axonal amplitude by
$\mathrm{TE}=60$\,ms --- both walls reproduced to ${\lesssim}0.2$\,pp. (The boundary-local-time
estimator is resolved by sub-stepping the walk to the extra-axonal pore scale, so it is
converged for both compartments; Sec.~\ref{sec:methods:mc}.) This pack-level agreement is an
internal consistency check --- Monte Carlo and closed form both evaluate the motional-narrowing
rate over the same realised geometry, so it confirms bookkeeping rather than the physics. The genuinely independent
validation is against the \emph{exact} lowest Robin eigenvalue on a single cylinder (the root
of $(kR)J_1/J_0=\rho R/D$, not merely its leading $\rho\,(2/R)$ limit;
App.~\ref{sec:app:estimator}): the wall-counting estimator lands on it to $<1\%$ across radii,
down to the sub-micron calibres that matter below --- so the rate the next section carries
across the calibre distribution is validated at the small-diameter end, not extrapolated. The central
physical fact follows from first principles: every wall encounter the Novikov--Burcaw
theory counts toward the time-dependent diffusion $D(t)$ is also a surface-relaxation
event, and the resulting rate $\rho\,(S/V)\approx5$\,s$^{-1}$ at the operating point
enters the transverse decay even though no gradient is present to encode the displacement.

The same physics shows directly in the multi-echo signal. Walking the \emph{full}
CPMG train with the forward Monte Carlo --- per-compartment bulk relaxation plus
first-principles wall relaxivity, with and without surface relaxivity, holding the
intrinsic bulk $T_2$ and the myelin volume fixed (Sec.~\ref{sec:methods:engine}) ---
the surface term visibly speeds the decay (Fig.~\ref{fig:signal}B), and more for the
fine, high-$S/V$ substrate than for the coarse one, even though both share the same
bulk $T_2$. Fed to the standard NNLS $T_2$-spectrum analysis, the apparent long-$T_2$
intra/extra (IE) pool shifts toward the short-$T_2$ myelin window (Fig.~\ref{fig:signal}C):
surface relaxivity shortens the apparent IE $T_2$, the more so at high $S/V$. In the raw walk
the \emph{integrated} short-$T_2$ fraction barely moves --- the sub-micron axons whose water
actually crosses the window lie below the simulator's calibre floor --- so the quantitative
myelin-water bias is carried by the full-distribution forward model of the next section, of
which this is the estimator-free, spectral-shift counterpart. The apparent $T_2$ of the intra/extra water is thus
$1/(1/T_{2,\mathrm{b}}+\rho\,S/V)$, with the geometric term inseparable from the
material one (Sec.~\ref{sec:theory:inseparability}); because $S/V=4/d$ for the
interior wall and scales likewise for the exterior, the shift is set by axon calibre
and packing --- the dependence the next section carries into the myelin water
fraction.

\subsection{Surface relaxivity biases the intra-axonal signal fraction}
\label{sec:results:fintra}

\begin{figure}[t]
  \centering
  \includegraphics[width=0.92\linewidth]{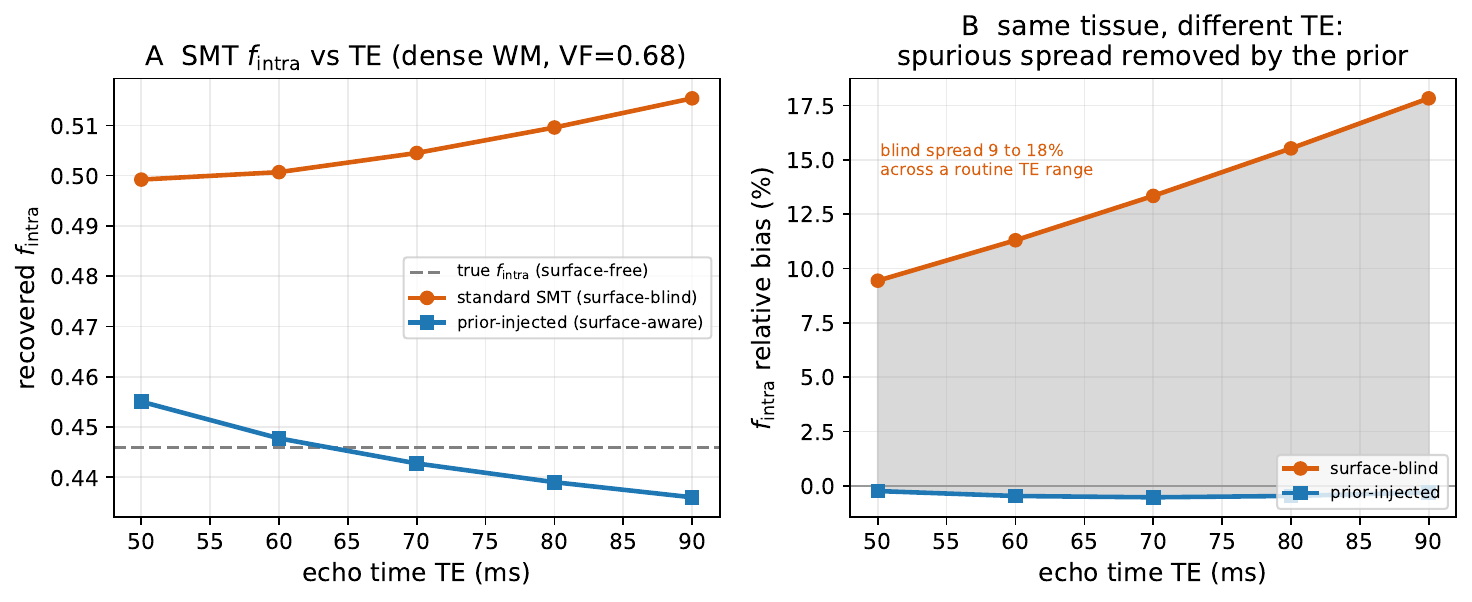}
  \caption{\textbf{A standard SMT fit returns the surface-biased $f_{\mathrm{intra}}$, and a
  histology prior corrects it at a single TE.} Dense WM ($\mathrm{VF}=0.68$), analytical WM
  forward with surface relaxivity on, fit by the standard spherical-mean technique in
  \texttt{dmipy-fit}.
  \textbf{(A)} Recovered $f_{\mathrm{intra}}$ vs echo time: surface-blind SMT (orange) drifts
  well above the surface-free truth (dashed); injecting the histology prior (blue) recovers it.
  \textbf{(B)} The same tissue read at different TE gives a spurious $f_{\mathrm{intra}}$ spread
  of ${\approx}15$--$30\%$ under a blind fit (a multi-site confound), removed by the prior.}
  \label{fig:smt}
  {\footnotesize\textit{Reproducible from} \texttt{figures/fig\_smt\_fintra.py}.}
\end{figure}

The most consequential reading of that $T_2$ shift is on the diffusion side. Every diffusion
microstructure model --- NODDI, the spherical-mean technique, the standard model --- estimates
an intra-axonal signal fraction $f_{\mathrm{intra}}$ and normalises the diffusion-weighted
signal by a $b{=}0$ acquired at the \emph{same} echo time, assuming the compartments share one
$T_2$. They do not: the interior and exterior walls carry different $S/V$, so surface relaxivity
gives the intra- and extra-axonal water different apparent $T_2$. To leading order the fraction
an unbiased estimator returns is the TE-weighted signal fraction
$w_c(\mathrm{TE})=f_c e^{-\mathrm{TE}R_c}/\sum_{c'}f_{c'}e^{-\mathrm{TE}R_{c'}}$
(Eq.~\eqref{eq:te_weights}), so the recovered $f_{\mathrm{intra}}$ inherits the surface
reweighting to leading order. That fractions are TE-dependent when compartment $T_2$ differ is
established \citep{Veraart2018}; what is new here is the \emph{source} we assign to the
compartmental $T_2$ difference --- a surface-relaxivity rate $\rho\,(S/V)$ tied to axon calibre
and packing --- and the resulting closed-form \emph{sign law}, below. We quantify the effect as
the bias in the signal weight $w_{\mathrm{intra}}$; this is the leading-order term that
propagates to any fraction estimator. This signal weight is distinct from the \emph{geometric}
fibre volume fraction $VF$ that sets $S/V$: it is a $b{=}0$-normalised, $T_2$-weighted water
fraction, related to $VF$ only through the compartment geometry and relaxation, and it is exactly
the quantity surface relaxivity biases (which is why the correction injects $VF$ as a prior rather
than reading it off the fit). It is not a strict upper bound on the recovered-parameter
bias: a fit with the diffusivities constrained returns the weight essentially directly,
and estimator model-mismatch can even \emph{amplify} it --- our spherical-mean fit below recovers
a bias of the same order as, and at points exceeding, the signal-weight estimate. We therefore
treat $w_{\mathrm{intra}}$ as the representative first-order magnitude, and confirm it with an
actual estimator (Fig.~\ref{fig:smt}).

\begin{figure}[t]
  \centering
  \includegraphics[width=\linewidth]{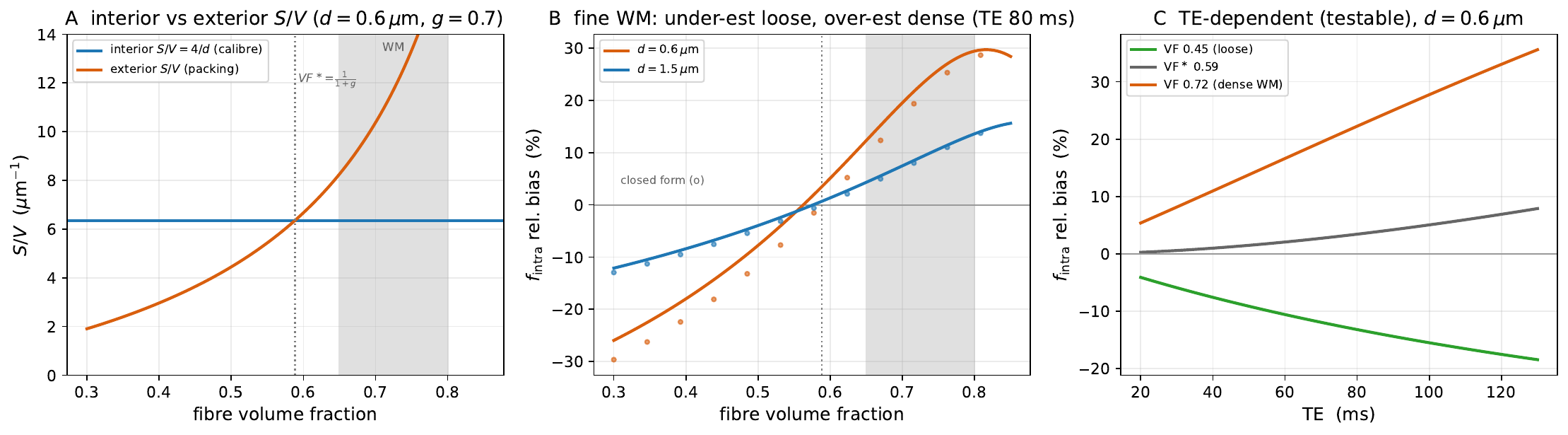}
  \caption{\textbf{Surface relaxivity biases the intra-axonal signal fraction, with a sign set
  by packing.}
  \textbf{(A)} Interior $S/V$ (calibre-set, flat) vs exterior $S/V$ (packing-set, rising) vs
  fibre volume fraction; they cross at $\mathrm{VF}^{\ast}=1/(1+g)$, and physiological WM (grey
  band) sits above it where the exterior wall dominates.
  \textbf{(B)} Intra-axonal signal-weight $w_{\mathrm{intra}}$ relative bias (the
  leading-order recovered-$f_{\mathrm{intra}}$ bias) vs fibre VF at clinical
  $\mathrm{TE}=80$\,ms (lines: exact distributed calculation; circles: closed form
  Eq.~\eqref{eq:fbias}): under-estimated when loose, over-estimated when dense, zero at
  $\mathrm{VF}^{\ast}$; ${+}12\%$ (relative) at $\mathrm{VF}=0.65$, rising to ${+}22\%$ at $0.72$.
  \textbf{(C)} The bias grows with echo time --- the testable fingerprint --- for loose,
  crossover, and dense packing.}
  \label{fig:fintra}
  {\footnotesize\textit{Reproducible from} \texttt{figures/fig\_fintra\_bias.py}.}
\end{figure}

The interior surface-to-volume is the per-axon $4/d$ (area/spin-weighted
$\langle 4/d\rangle_V=2\langle r\rangle/\langle r^2\rangle$, set by calibre alone); the exterior
is set by packing --- total outer-wall perimeter over extra-axonal area. Their ratio
(Eq.~\eqref{eq:svratio_theory}) reduces to the purely geometric $(1-\mathrm{VF})/(g\,\mathrm{VF})$,
\emph{independent of the calibre distribution} (which sets only the absolute scale), and the two
are equal at a crossover packing $\mathrm{VF}^{\ast}=1/(1+g)\approx0.59$ (Fig.~\ref{fig:fintra}A):
below it the interior wall dominates and intra water relaxes faster; above it the extra-axonal
water, squeezed into thin high-$S/V$ crevices between tightly packed fibres, relaxes faster.

Carried into the fraction estimate, the intra/extra \emph{odds} are rescaled by a single
exponential,
\begin{equation}
  \frac{f_{\mathrm{intra}}^{\mathrm{app}}}{1-f_{\mathrm{intra}}^{\mathrm{app}}}
  = \frac{f_{\mathrm{intra}}}{1-f_{\mathrm{intra}}}\,
    \exp\!\big[-\mathrm{TE}\,\rho\,\big((S/V)_{\mathrm{int}}-(S/V)_{\mathrm{ext}}\big)\big],
  \label{eq:fbias}
\end{equation}
so $f_{\mathrm{intra}}$ is \emph{under}-estimated in loosely packed white matter and
\emph{over}-estimated in densely packed white matter, crossing zero at $\mathrm{VF}^{\ast}$
(Fig.~\ref{fig:fintra}B; the closed form of Eq.~\eqref{eq:fbias} reproduces the sign law and
crossover exactly and tracks the magnitude of the exact distributed-calibre calculation to
${\lesssim}4$\,pp, slightly \emph{under}-estimating it --- by Jensen, the distributed thin-calibre
tail leaves the surviving intra water higher than the mean-rate approximation). Physiological
white matter (fibre
$\mathrm{VF}\approx0.65$--$0.8$) sits \emph{above} the crossover, where at clinical PGSE
($\mathrm{TE}=80$\,ms) and the cited $\rho$ the intra-axonal signal weight is over-estimated by
${\approx}12\%$ at $\mathrm{VF}=0.65$, rising to ${\approx}22\%$ by $\mathrm{VF}=0.72$. At the
denser end the exterior $S/V$ approaches ${\sim}11\,\mu$m$^{-1}$ (sub-$0.1\,\mu$m crevices) and
the mean-field plateau of Eq.~\eqref{eq:svratio_theory} becomes marginal (extra-axonal water may
not motionally average over the local $S/V$; Sec.~\ref{sec:discussion}), so the largest values
are an upper edge, not a central estimate --- and, since the local-$S/V$ distribution can only
raise the exterior rate (Jensen), the robust ${\approx}12\%$ at $\mathrm{VF}\,0.65$ is a
\emph{lower} bound on the dense side. For traceability:
at $\mathrm{VF}=0.65$ the interior and exterior densities are $\svin\!\approx\!6.3$ and
$S_{\mathrm{ext}}/V_{\mathrm{ext}}\!\approx\!8.3\,\mu$m$^{-1}$, a differential surface rate
$\rho\,(S_{\mathrm{ext}}/V-\svin)\!\approx\!2.3\,$s$^{-1}$ that over $\mathrm{TE}=80$\,ms shifts
the intra/extra odds by ${\approx}20\%$ --- a ${\approx}{+}9\%$ mean-rate fraction bias, which
the exact calibre-distributed calculation (Fig.~\ref{fig:fintra}B) raises to ${+}12\%$ (the
${\lesssim}3$\,pp closed-vs-exact gap, from the thin-calibre tail). All $f_{\mathrm{intra}}$
percentages here are \emph{relative} fraction biases.

The exponent of Eq.~\eqref{eq:fbias} is \emph{linear} in $\rho$, so the same order-of-magnitude
uncertainty that bounds the MWF result below applies here directly and symmetrically: the
${\approx}12\%$ is quoted at the cited $\rho$ and scales roughly linearly with it (the
log-odds shift doubles as $\rho$ doubles). Unlike the MWF term below --- which we show rides
${\sim}50\times$ beneath \emph{single-voxel} noise --- this is a first-order bias on a headline
metric; being deterministic in the substrate it is a \emph{spatially structured systematic},
not zero-mean noise, so it does not average down as $1/\sqrt{N}$ but shifts a regional or cohort
mean by close to its full value. Whether it clears single-voxel diffusion-fit scatter depends on
the protocol and is not established here; its signature is regional and cross-cohort, not
per-voxel.

The bias carries a \emph{testable} signature: it grows monotonically with echo time
(Fig.~\ref{fig:fintra}C), so an $f_{\mathrm{intra}}$ that drifts with TE at fixed $b$ is
consistent with it. This is necessary but not sufficient evidence --- the same TE drift arises
from any compartmental $T_2$ difference (intrinsic intra/extra bulk $T_2$, CSF partial volume,
exchange), the generic TEdDI effect \citep{Veraart2018}. What would distinguish the surface term
is its \emph{packing dependence}: the sign and magnitude track fibre volume fraction through
Eq.~\eqref{eq:svratio_theory}, reversing below $\mathrm{VF}^{\ast}$, whereas an intrinsic-$T_2$
offset does not.

So far the bias is the reweighting of the intra-axonal \emph{signal weight}; we confirm it
survives an actual estimator and that a prior corrects it (Fig.~\ref{fig:smt}). We forward-simulate
a dense-WM multi-shell signal with surface relaxivity on, and fit it with the standard
spherical-mean technique (SMT: intra stick $+$ extra zeppelin) \citep{Kaden2016}. Because the
surface factor is $b$-independent it propagates through the whole multi-shell fit, so the
recovered $f_{\mathrm{intra}}$ inherits the bias and \emph{drifts with echo time} --- from
${\approx}{+}15\%$ at $\mathrm{TE}=50$\,ms to ${\approx}{+}30\%$ at $90$\,ms (the surface term
isolated by differencing against a surface-off fit, whose recovered fraction is TE-independent
to ${<}0.01$, so the drift is the surface term and not a fitting artefact;
Sec.~\ref{sec:methods:smt}). This confirms the effect is not merely a $b{=}0$ prefactor: it
propagates to what a real estimator returns, at a magnitude comparable to (here somewhat above)
the analytical signal-weight bias. A cohort of healthy controls scanned at different echo
times --- routine across sites --- would thus report spuriously different $f_{\mathrm{intra}}$
for identical tissue. The degeneracy that forbids \emph{estimating} the surface term at one TE
(Sec.~\ref{sec:theory:inseparability}) does not forbid \emph{correcting} for it: injecting a
region's histology-derived calibre prior (so that
$\rho\,[(S/V)_{\mathrm{int}}-(S/V)_{\mathrm{ext}}]$ is known, as NODDI already fixes
diffusivities) undoes the odds rescaling of Eq.~\eqref{eq:fbias} at a single TE, collapsing the
$15$--$30\%$ drift to a ${\sim}6\%$ residual (Fig.~\ref{fig:smt}) and harmonising the cohort.

\subsection{Surface relaxivity also biases the recovered myelin water fraction}
\label{sec:results:mwf}

The same $S/V$ difference reads through relaxometry too, in the myelin water fraction --- a
smaller, structural companion that ties the diffusion and relaxometry pictures to one physics.
Carrying that surface rate into the standard myelin-water analysis at \emph{fixed
myelin volume} (Sec.~\ref{sec:methods:mwf}), the signal-weighted IE apparent $T_2$
swings from $68.2$\,ms for large-calibre ($3\,\mu$m outer) fibres to $44.9$\,ms for fine
($0.45\,\mu$m outer) fibres --- a ${\sim}34\%$ change driven entirely by the surface term
$\rho\,(S/V)$, with no change in intrinsic tissue $T_2$ and none in myelin content.
This $T_2$ shift is NNLS-independent (Eq.~\eqref{eq:r2app} read directly) and is the
paper's most robust quantitative statement.

\begin{figure}[t]
  \centering
  \includegraphics[width=\linewidth]{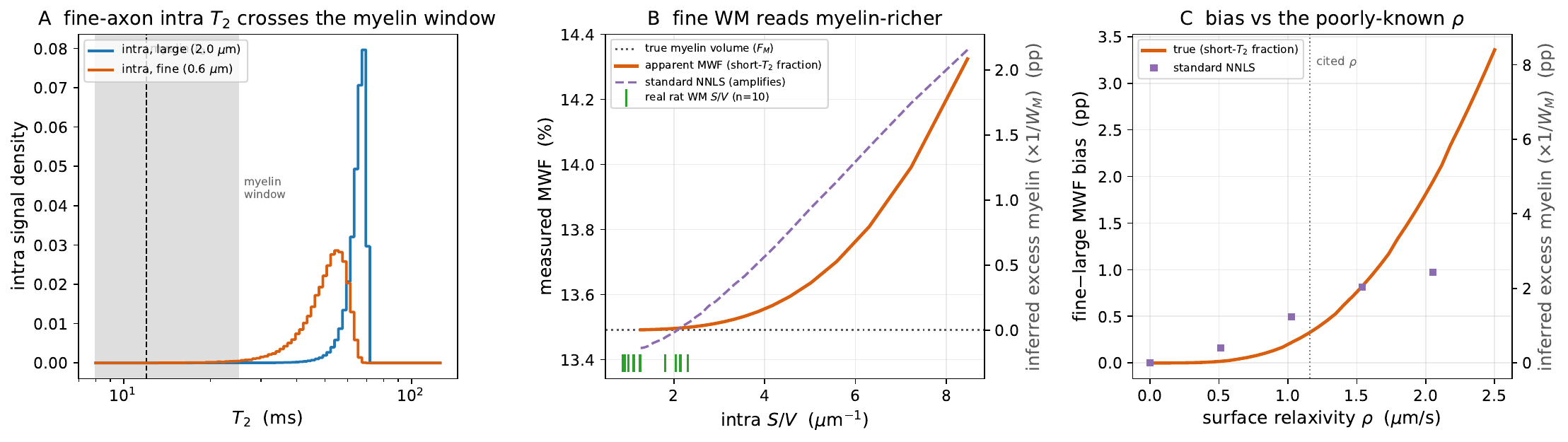}
  \caption{\textbf{Surface relaxivity biases the recovered myelin water fraction: fine
  white matter reads myelin-richer.}
  \textbf{(A)} Mechanism. Each axon relaxes at $1/T_{2,\mathrm b}+\rho\,(4/d)$, so the
  intra-axonal $T_2$ is a distribution over calibre; for fine axons ($0.6\,\mu$m, orange)
  it shifts down and grows a tail that crosses \emph{below} the $25$\,ms myelin window
  (shaded), where it is counted as myelin water --- the large-calibre pool (blue) does
  not.
  \textbf{(B)} At fixed myelin volume, the apparent MWF (short-$T_2$ fraction, grid-free)
  rises above the true myelin fraction $F_M$ (dotted) as $S/V$ increases --- fine WM reads
  myelin-richer; standard NNLS (dashed) amplifies it. Right axis: inferred excess myelin
  content ($\times 1/W_M$), the $2.5\times$-amplified reading. Green rug: interior $S/V$ of
  $10$ real rat cross-sections (AxonDeepSeg; Sec.~\ref{sec:methods:real}).
  \textbf{(C)} The bound: fine- ($0.6\,\mu$m outer) vs large-calibre ($2.0\,\mu$m outer) MWF
  excess vs the poorly-known $\rho$ --- $0.33$\,pp (${\approx}0.82$\,pp myelin content) at the
  cited $\rho$, climbing super-linearly to $3.4$\,pp (${\approx}8.4$\,pp myelin) at
  $2.5\,\mu$m/s.}
  \label{fig:mwf}
  {\footnotesize\textit{Reproducible from} \texttt{figures/fig\_mwf\_sv\_bias.py}.}
\end{figure}

\begin{figure}[t]
  \centering
  \includegraphics[width=0.62\linewidth]{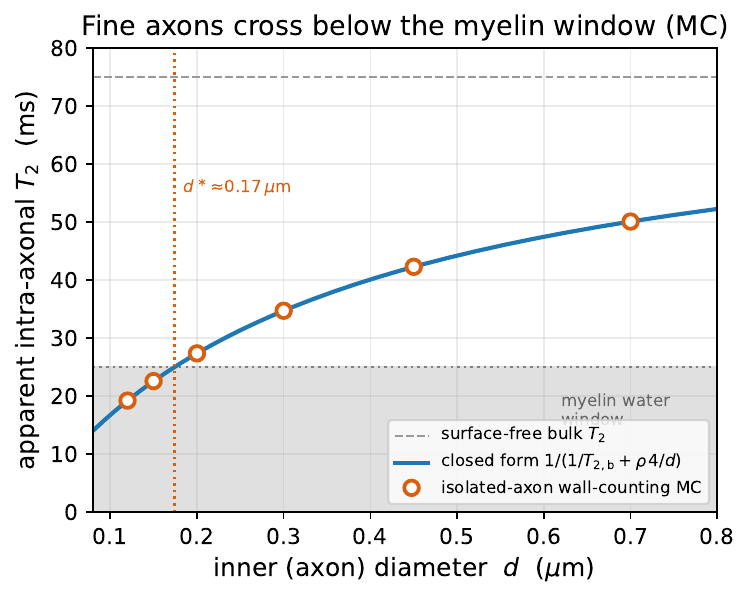}
  \caption{\textbf{Fine axons cross below the myelin window --- shown directly in the
  Monte Carlo.} Gradient-free wall-counting walks in \emph{isolated} reflecting cylinders
  (confined interior, where the estimator is exact) recover the apparent intra-axonal $T_2$
  (circles) at a range of inner calibres. They sit on the closed form
  $1/(1/T_{2,\mathrm b}+\rho\,4/d)$ (blue) to the digit and cross below the $25$\,ms myelin
  window (shaded) at $d^{\ast}\approx0.17\,\mu$m, confirming the mechanism the distributed
  forward model integrates over the calibre distribution. The packed CPMG cannot reach this
  sub-micron tail (step cost $\propto1/R^2$); the interior effect needs only the single-axon
  walk.}
  \label{fig:crossing}
  {\footnotesize\textit{Reproducible from} \texttt{figures/fig\_crossing\_mc.py}.}
\end{figure}

That shift has a specific, and at first counter-intuitive, consequence for the recovered
MWF. The intra-axonal water is not one $T_2$ but a distribution over calibre: each axon of
\emph{inner} (axon) diameter $d_{\mathrm i}=g\,d$ carries an interior $S/V=4/d_{\mathrm i}$, so its
water relaxes at $1/T_{2,\mathrm b}+\rho\,(4/d_{\mathrm i})$, and the \emph{thinnest} axons are
shortened most. Once the inner diameter falls below a critical calibre
$d^{\ast}=4\rho\,/\,(1/T_2^{\mathrm{cut}}-1/T_{2,\mathrm b})$ --- $0.17\,\mu$m (an inner/axon
diameter) at the cited $\rho$ --- that axon's water crosses \emph{below} the $25$\,ms myelin
window and is counted, by the window's own definition, as myelin water (Fig.~\ref{fig:mwf}A). Fine
white matter therefore reads myelin-\emph{richer}, not poorer: the short-$T_2$ signal
fraction, computed grid-free with no NNLS, rises above the true myelin fraction of $13.49\%$
(recovered exactly at large calibre) to ${\approx}14.3\%$ at the finest calibre swept
(Fig.~\ref{fig:mwf}B); between the reference calibres of the bound below the fine ($0.6\,\mu$m
outer) vs large ($2.0\,\mu$m outer) excess is ${\approx}0.33$\,pp at the cited $\rho$.
The standard chi-square-regularised NNLS estimator reproduces the same sign and amplifies
it ($13.44\!\to\!14.35\%$; dashed), as it also assigns near-window intra water to the
short-$T_2$ pool. The crossing is not merely a forward-model construction. Walking \emph{isolated} confined
axons across a range of inner calibres with the wall-counting Monte Carlo --- which is exact
for the confined interior (its single-cylinder rate matches the closed form $\rho\,(4/d)$ to
the digit and the exact Robin eigenvalue to $<1\%$, App.~\ref{sec:app:estimator}) --- recovers
an apparent intra-axonal $T_2=1/(1/T_{2,\mathrm b}+\rho\,4/d_{\mathrm i})$ that falls from
$27.4$\,ms at $d_{\mathrm i}=0.20\,\mu$m to $22.6$\,ms at $0.15\,\mu$m and $19.2$\,ms at
$0.12\,\mu$m --- crossing \emph{below} the $25$\,ms window at $d^{\ast}\approx0.17\,\mu$m
(Fig.~\ref{fig:crossing}). What the from-scratch simulator cannot yet integrate is a
\emph{packed} substrate: a dense pack's calibre floor (step cost $\propto 1/R^2$) truncates the
sub-micron tail, so the packed CPMG shows the spectral shift but a negligible integrated MWF
drift. The crossing is an interior effect, however, so the single-axon walk establishes it, and
the distributed forward model then carries it across the calibre distribution.

Read as myelin \emph{content} --- MWF's purpose --- the effect is larger than the
water-fraction bias. The contaminating signal is intra-axonal water (${\approx}100\%$
water) misassigned to the myelin pool, whose water content is only $W_M{\approx}0.4$
(${\approx}60\%$ dry lipid and protein) \citep{West2018}; converting apparent MWF to
myelin volume at that ratio over-states myelin by $1/W_M{\approx}2.5\times$ the
water-fraction bias. The $0.33$\,pp fine-vs-large MWF excess at the cited $\rho$ thus
implies a ${\approx}0.82$\,pp over-estimate of myelin content between the two regions at
equal true myelin (Fig.~\ref{fig:mwf}B, right axis).

The magnitude is contingent on $\rho$, which for myelinated tissue is not independently
established --- the cited $\rho=1.16\,\mu$m/s is a coefficient of a relaxation--diffusion
model fitted to histology \citep{Barakovic2023}, not a measured axolemma relaxivity, and
is plausibly uncertain by an order of magnitude. Because $d^{\ast}$ grows linearly with
$\rho$ while the volume-weighted fraction of axons below it grows super-linearly, the
effect climbs steeply: the fine-vs-large MWF excess rises from $0.33$\,pp at the cited
$\rho$ to $3.36$\,pp at $2.5\,\mu$m/s --- equivalently, $0.82\!\to\!8.4$\,pp of inferred
myelin content (Fig.~\ref{fig:mwf}C), no longer negligible at the upper end and a fraction
of the $2$--$5$\,pp MWF change attributed to demyelination
\citep{Laule2004, Laule2006, Kitzler2012}. The structural claim ($S/V$ and $\rho$
unobserved, the term inseparable) does not depend on the value. Being deterministic in the
substrate the bias does not average away, but at clinical SNR it rides far beneath the
\emph{single-voxel} noise: at $\mathrm{SNR}=150$ a single substrate's recovered MWF scatters by
${\sim}4.7$\,pp, ${\sim}14\times$ the cited-$\rho$ systematic. That factor compares a single
voxel's absolute-MWF scatter against the cross-region excess and so overstates how buried the
\emph{contrast} is --- some of the NNLS ill-conditioning and realisation variance cancels in a
matched fine-vs-large difference. The systematic is nonetheless exposed only by averaging: since
it is spatially structured (it tracks calibre and packing) rather than zero-mean, it survives
averaging that noise does not, but exposing a ${\sim}0.33$\,pp offset above a ${\sim}4.7$\,pp
single-voxel floor still needs a region of order $(4.7/0.33)^2\!\sim\!200$ voxels even in the
favourable uncorrelated-noise case. It is thus a small, cohort- or ROI-level systematic, not a
per-voxel readout.

\subsection{The bias as a tract demyelinates}
\label{sec:results:demyel}

\begin{figure}[t]
  \centering
  \includegraphics[width=\linewidth]{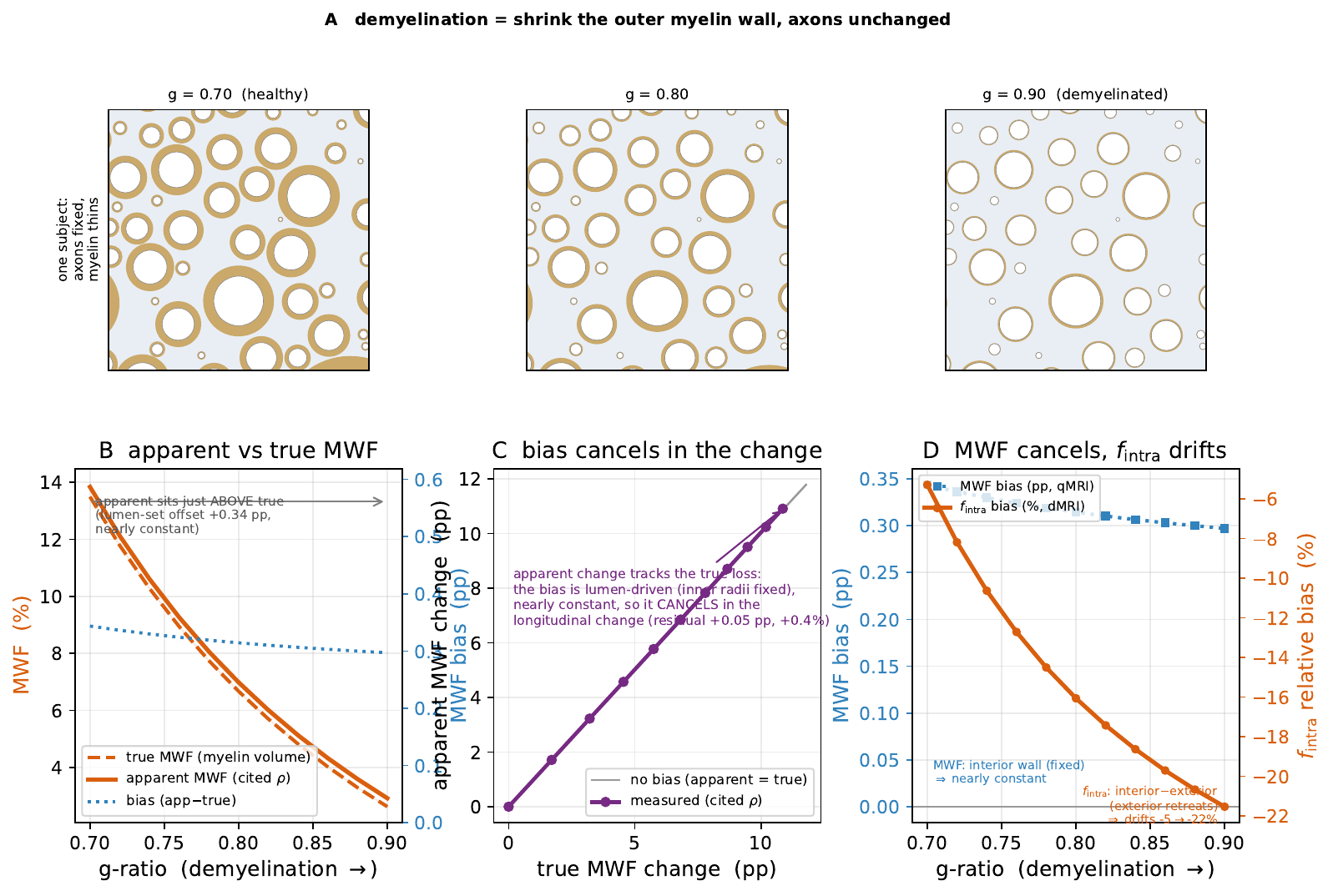}
  \caption{\textbf{As a tract demyelinates, the surface-relaxivity bias cancels in the
  longitudinal change.} Within-subject trajectory ($24$ packs; axons frozen, myelin thinned
  from $g_0=0.70$ to $g=0.90$; Sec.~\ref{sec:methods:demyel}), same forward model and
  MC-validated rate as Fig.~\ref{fig:mwf}.
  \textbf{(A)} One subject's substrate at three stages: axon centres and inner radii frozen,
  the outer myelin wall $r_{\mathrm{out}}=r_{\mathrm{in}}/g$ shrinking.
  \textbf{(B)} Apparent (solid, cited $\rho$) sits just above true (dashed) MWF, and both
  fall steeply; the bias (dotted, right axis) is small and nearly constant because it is set
  by the fixed inner radii, not the retreating outer wall.
  \textbf{(C)} Apparent vs true MWF \emph{change}: the measured curve lies on the no-bias
  diagonal --- the lumen-driven offset cancels, so the apparent loss tracks the true loss to
  within ${\sim}0.4\%$.
  \textbf{(D)} The unifying contrast: along the same trajectory the relaxometry MWF bias (blue,
  left axis) stays flat and cancels --- it reads the fixed \emph{interior} wall --- while the
  diffusion $f_{\mathrm{intra}}$ bias (orange, right axis) \emph{drifts} from ${\approx}{-}5\%$
  to ${\approx}{-}21\%$, because it reads the interior$-$exterior differential and the exterior
  wall retreats as myelin thins.}
  \label{fig:demyel}
  {\footnotesize\textit{Reproducible from} \texttt{figures/fig\_demyelination\_bias.py}.}
\end{figure}

The bias so far is a cross-region contrast at fixed myelin. The clinically central
question is longitudinal: as a tract demyelinates, does the surface term make the apparent
MWF \emph{change} mis-state the true myelin loss? We trace this within-subject (24 realised
packs; axons frozen, myelin thinned from $g_0=0.70$ toward $g=0.90$;
Sec.~\ref{sec:methods:demyel}), one subject's cross-section at three stages shown in
Fig.~\ref{fig:demyel}A --- the axon centres and inner (lumen) radii frozen, only the outer
myelin wall retreating.

The key is that the crossing which causes the bias is set by the \emph{inner} diameter, and
primary demyelination preserves the lumen. The bias is therefore a nearly \emph{constant}
offset: apparent MWF sits just above true throughout, by $+0.34$\,pp at baseline and
$+0.30$\,pp at $g=0.90$ (Fig.~\ref{fig:demyel}B) --- the small drift coming only from the
water-fraction normalisation as $40\%$-water myelin converts to $100\%$-water extra-axonal
space, not from the crossing itself. Because a within-subject \emph{change} subtracts the
baseline, this near-constant offset cancels: over a trajectory in which the true MWF falls
$13.49\!\to\!2.63\%$ (a $10.86$\,pp drop) the apparent MWF falls $10.91$\,pp, an
over-report of ${\sim}0.05$\,pp --- of order $0.4\%$ of the true change, and within the forward
model's own idealisation (Fig.~\ref{fig:demyel}C). So the
surface-relaxivity confound is a \emph{cross-region} effect, not a longitudinal one: for
primary demyelination (a g-ratio rise at preserved axons) it very nearly cancels in
same-substrate tracking, because it is tied to the axon lumen the pathology leaves intact.
Axonal loss or a shift in the fine-calibre population would break this cancellation ---
they alter the very inner-diameter distribution the offset is set by --- so the statement is
bounded to lumen-preserving demyelination.

Crucially, the \emph{diffusion} intra-axonal fraction behaves oppositely along the same
trajectory, and the contrast unifies the two halves of this paper through the g-ratio. The MWF
bias is set by the \emph{interior} crossing (the inner axon wall, which demyelination
preserves), so it is constant and cancels. The $f_{\mathrm{intra}}$ bias of Eq.~\eqref{eq:fbias}
is instead set by the interior-\emph{minus}-exterior differential $\rho[(S/V)_{\mathrm{int}}-(S/V)_{\mathrm{ext}}]$,
and the g-ratio enters both walls: the exterior wall is the fibre boundary at $d_{\mathrm i}/g$,
so as myelin thins it \emph{retreats} and $(S/V)_{\mathrm{ext}}$ collapses (here $5.4\!\to\!2.9\,\mu$m$^{-1}$
over $g\,0.70\!\to\!0.90$) while $(S/V)_{\mathrm{int}}$ stays fixed. The differential therefore
\emph{grows}, and the $f_{\mathrm{intra}}$ bias \emph{drifts} rather than cancels --- from
${\approx}{-}5\%$ at baseline to ${\approx}{-}21\%$ at $g=0.90$ (Fig.~\ref{fig:demyel}D), at a
clinical $\mathrm{TE}=80$\,ms. (This trajectory's baseline sits at fibre $\mathrm{VF}=0.55$,
just \emph{below} the crossover $\mathrm{VF}^{\ast}(g_0)=0.59$, so its $f_{\mathrm{intra}}$ bias
is \emph{negative} --- interior-dominated, opposite in sign to the dense-WM headline of
Sec.~\ref{sec:results:fintra} --- and grows more negative as the exterior wall retreats.)
So one $S/V$ physics gives longitudinally opposite readouts: the
relaxometry MWF change is nearly confound-free, whereas a diffusion $f_{\mathrm{intra}}$ change
across a demyelinating tract carries a large, TE-dependent surface artefact --- and, because the
trajectory sweeps fibre VF across the crossover $\mathrm{VF}^{\ast}=1/(1+g)$, the sign of the
$f_{\mathrm{intra}}$ bias can even reverse mid-trajectory for a denser healthy baseline.

\section{Discussion}
\label{sec:discussion}

The transverse rate a multi-echo spin echo measures is bulk relaxation plus an isotropic,
TE-linear surface rate $\rho\,(S/V)$ inseparable from it, with $S/V$ unobserved. The
consequence for the myelin water fraction is specific: surface relaxivity shortens the
intra/extra apparent $T_2$, and for the thinnest axons it drags that $T_2$ below the
myelin window, where the water is counted as myelin --- so fine white matter reads
myelin-\emph{richer} at fixed myelin content, by a substrate-dependent amount (${\sim}0.1$
percentage points at the cited $\rho$, super-linearly more --- $>1$\,pp --- at the upper
end of its order-of-magnitude uncertainty). It is no artefact: a wall-counting Monte Carlo
reproduces the rate from first principles, and the short-$T_2$ crossing is a property of
the window's own definition, present before any estimator sees the signal.

\paragraph{The same physics from two communities.} The contribution is partly one of
perspective: the time-dependent extra-axonal $D(t)$ of randomly packed media is well
established in diffusion MRI \citep{Novikov2014, Burcaw2015, Fieremans2016}, and the same
wall collisions, seen from relaxometry, are surface-relaxation events (Sec.~\ref{sec:theory:dual}).
The corollary: the ``$T_2$'' a relaxometry experiment reports is contaminated by a
geometry-dependent surface rate with no gradient played.

\paragraph{Implications for myelin water imaging.} MWF is read as a myelin-content index
and compared across tracts, subjects, and time. Because per-voxel MWF is noisy, studies
average it over a region of interest --- a callosal subregion, a tract, lobar white
matter --- and compare ROI means; subtracting two such means is where the surface term
bites. Each ROI carries its own offset set by its own thin-axon population, so
\begin{equation}
  \Delta\mathrm{MWF}_{\mathrm{app}}
  = \underbrace{\Delta\mathrm{MWF}_{\mathrm{true}}}_{\text{myelin difference}}
  \;+\;\big[\,\beta_A-\beta_B\,\big],
  \qquad \beta\ =\ F_I\,\Phi(\rho),
  \label{eq:roi_diff}
\end{equation}
where $F_I$ is the intra-axonal signal fraction and $\Phi(\rho)$ the volume-weighted
fraction of that intra-axonal water dragged below the myelin window (the thin-calibre tail),
so $\beta$ is larger for the finer region. The second bracket ---
a pure calibre difference --- adds to the myelin difference and cannot be separated from it
within the multi-echo data. It vanishes only when the two ROIs share their fine-axon
population, and is largest where calibre differs systematically: between callosal
subregions (the finer genu and splenium versus the large-diameter mid-body
\citep{Aboitiz1992} --- predicting the finer genu/splenium read myelin-\emph{richer} than
the mid-body at equal myelin), across development or ageing, or between patient and control
groups when disease shifts the calibre spectrum. At ${\sim}0.1$\,pp (${\sim}0.24$\,pp of
inferred myelin content) at the cited $\rho$ it is a small fraction of a gross regional
contrast, but a meaningful one for the subtle, ${\sim}1$\,pp differences cross-sectional
myelin studies pursue --- and, being deterministic in the substrate, a cohort-level
systematic that shifts a group mean rather than averaging out. The comparison it spares is
the field's cleanest, the \emph{same region of the same subject over time}: there the offset
is a nearly-constant, lumen-set term that cancels in the change
(Sec.~\ref{sec:results:demyel}), because primary demyelination leaves the inner diameters
the crossing depends on intact. None of this invalidates MWF; it bounds a contribution of
the same $S/V$-driven family as the documented exchange bias \citep{Harkins2012}, not
previously quantified as a surface-relaxivity term inseparable from the bulk $T_2$.

\paragraph{What counts as myelin water.} The mechanism is not really about axons: the
short-$T_2$ window captures \emph{any} water whose $T_2$ is dragged below it by surface
relaxivity, which happens once $S/V$ exceeds
$(1/T_2^{\mathrm{cut}}-1/T_{2,\mathrm b})/\rho\approx 23\,\mu$m$^{-1}$ at the cited $\rho$
(a sub-micron scale) and $\approx 11\,\mu$m$^{-1}$ at the upper $\rho$. This is the very
physics that makes bilayer-trapped myelin water short-$T_2$ in the first place, so MWF has
always measured ``high-$S/V$ trapped water'', of which myelin is merely the dominant source
in health. It follows --- mechanism-predicted, not simulated here --- that \emph{any}
membrane-dense sub-micron compartment qualifies: myelin debris, fine reactive-glial
processes or microvacuolation could add short-$T_2$ signal counted as myelin and prop up
apparent MWF, while whole cells ($S/V\!<\!1\,\mu$m$^{-1}$) are far too large to cross. This
widens the caveat from a calibre confound to a statement about what the short-$T_2$ pool
contains, and is a natural target for a drop-in Monte Carlo of explicit pathological
microstructure.

\paragraph{Idealised packing, and where Monte Carlo becomes essential.} Our closed forms
are exact for the substrate we assume --- parallel, circular, non-touching cylinders packed
by random sequential adsorption --- and for that geometry they are also the \emph{efficient}
route. A Monte-Carlo pack pays a sub-step cost that scales as $1/R^2$ in its smallest axon
(an impermeable-wall relaxivity weight needs steps $\ll R$), so the very calibres that
dominate the bias are the most expensive to simulate \citep{Hall2009}, while the analytics
returns them for free. The idealisation has a specific blind spot,
though. The exterior rate uses the \emph{mean} $S/V$ --- total wall perimeter over total
extra-axonal area --- and a mean cannot see the \emph{distribution of local} $S/V$ that
realistic packing creates: where axons deform, touch, undulate or cross \citep{Xu2018},
extra-axonal water is trapped in thin interstitial crevices of very high local $S/V$, and
the window-crossing the intra-axonal calibre tail undergoes could occur there too ---
extra-axonal water read as myelin. Whether it does hinges on motional averaging, and two
\emph{distinct} length scales must not be conflated. A local sheet of high $S/V$ shortens the
$T_2$ enough to cross the myelin window once its width falls below
${\sim}90$\,nm ($S/V\gtrsim23\,\mu$m$^{-1}$ at the cited $\rho$) --- a scale the extra-axonal
space genuinely reaches. But whether that water \emph{averages} with the bulk is a
\emph{separate} condition, set by how fast a spin escapes the pocket: at a $90$\,nm width the
traversal time $w^2/D\!\sim\!4\times10^{-3}$\,ms is far below the echo spacing, so an
\emph{open} $90$\,nm sheet averages comfortably and feels only the mean $S/V$. Averaging fails
only where the pocket is kinetically \emph{isolated} --- narrow throats or near-contacts whose
bottleneck traversal time approaches the echo spacing --- a percolation-scale property of the
packing, not of the sheet width, and one we do not quantify here. So the mean-field rate is a
lower bound (Jensen; Sec.~\ref{sec:theory:closedforms}) and the size of the correction is set by
this unquantified isolation scale. This is the regime the analytics cannot reach and only a
Monte Carlo of non-idealised geometry can resolve --- and, by the same $1/R^2$ argument, the
most costly to run. The reading is therefore a duality, not a ranking: the closed form should carry the
idealised packing, where Monte Carlo is wasteful, and Monte Carlo should be reserved for the
irreducibly non-idealised part --- crevices, undulation, crossings --- where the closed form
has nothing to say.

\paragraph{Why the correction injects the fraction rather than fitting it.} The quantity a
diffusion model returns is the intra-axonal \emph{signal} fraction, not the geometric fibre
volume fraction $VF$; a spherical-mean fit even ties its extra-axonal (tortuosity) diffusivity to
that signal fraction, so the packing geometry that sets $S/V$ is never an independent output.
Because the surface term is $b$-independent and inseparable from the bulk $T_2$, $S/V$ --- and
hence $VF$ --- is not identifiable from the fit, and feeding the fitted fraction back into the
surface term would be circular (it is the very quantity the surface term biases). The correction
therefore treats $VF$ as a \emph{bounded prior} supplied from regional histology --- exactly the
ingredient a cross-TE cohort harmonisation needs and can plausibly obtain --- and leaves the fit
to return only the corrected fraction.

\paragraph{What could resolve it, and what cannot.} The degeneracy is structural, so no
refinement of the multi-echo acquisition alone breaks it: echo spacing, train length and
signal-to-noise change precision, not the non-identifiability. Diffusion is, in
principle, the orthogonal axis --- the same wall collisions, read with a gradient. The
elegant point is that the disorder enters diffusion through the \emph{same} fibre
orientation distribution $\mathcal F(\hat{\mathbf n})$ as the signal convolution
Eq.~\eqref{eq:fod_conv}: in the low-$b$, short-time regime the apparent diffusivity along
$\hat{\mathbf g}$ is that FOD convolved with a disorder kernel,
\begin{equation}
  \frac{D(t;\hat{\mathbf g})}{D_0}=\int_{S^2}\mathcal F(\hat{\mathbf n})\,
  \kappa_{\mathrm{dis}}(\hat{\mathbf g}\!\cdot\!\hat{\mathbf n};t)\,\mathrm d\hat{\mathbf n},
  \quad
  \kappa_{\mathrm{dis}}=1-\tfrac{4}{3\sqrt\pi}\sqrt{D_0 t}\,
  \Big[\textstyle\sum_c f_c\,(S/V)_c\Big]\big[1-(\hat{\mathbf g}\!\cdot\!\hat{\mathbf n})^2\big],
  \label{eq:kdis}
\end{equation}
with $(S/V)_c$ the interior $\svin$ or exterior $S_{\mathrm{ext}}/V_{\mathrm{ext}}$. Because
$1-(\hat{\mathbf g}\!\cdot\!\hat{\mathbf n})^2=\tfrac23[1-P_2(\hat{\mathbf g}\!\cdot\!\hat{\mathbf n})]$
carries only $\ell=0$ and $\ell=2$, the surface disorder is a \emph{rank-2} function on the
sphere --- round ($\ell{=}0$) to relaxation, fibre-shaped ($\ell{=}0{+}2$) to diffusion --- and
spherical convolution preserves that band limit for \emph{any} FOD. The orientation-invariant
total $S/V$ is then the $\ell=0$ (trace) part, which an isotropic (trace) oscillating-gradient
acquisition or a spectrally tuned $b$-tensor \citep{Lundell2019, Westin2016} recovers
independent of fibre dispersion. In practice it is gradient-limited: the short-time regime needs
OGSE frequencies whose deliverable $b$-value collapses as $\sim$$1/f^3$ \citep{Does2003} ---
sitting at the corner frequency of a micron axon needs $G\!\sim\!70$--$260\,$T/m, one to two
orders beyond even preclinical inserts --- so the fine calibres that carry the effect lie out of
reach, and a realisable protocol is left to future work. What no route removes
is the absolute calibration: the inner- and outer-wall relaxivities
$\rho_{\mathrm i},\rho_{\mathrm e}$ --- which may differ, and scale the bias --- are not
independently established for myelinated tissue, and the myelin-water pool itself, with its
very large bilayer $S/V$, is plausibly surface-dominated in a way we have not modelled
here.

\paragraph{Relation to exchange and other MWF confounds.} MWF is already known to
depend on factors other than myelin content: $B_1$/flip-angle imperfections, which
the extended-phase-graph fit corrects \citep{Prasloski2012, Lebel2010}; iron and
susceptibility; and inter-compartment water \emph{exchange}, which Harkins, Dula and
Does showed makes the apparent MWF depend on axon size \citep{Harkins2012, Dula2010}.
The exchange bias is, like ours, driven by surface-to-volume ratio, and both are largest
at high $S/V$. They are \emph{mechanistically distinct} --- exchange moves magnetisation
between pools of different $T_2$, whereas surface relaxivity destroys coherence at the wall
and, for the thinnest axons, moves intra water across the window --- but \emph{operationally
entangled}: both act at the small-calibre end and neither is separable within the multi-echo
data. We deliberately isolate surface relaxivity by making the myelin impermeable in the
simulation, so the rate validated in Fig.~\ref{fig:signal}A and the bias of
Fig.~\ref{fig:mwf} are exchange-free; we do not claim to have measured their \emph{relative}
magnitude, which would require permeable walls and is left to future work. Myelin is a strong
barrier (slow apparent exchange, $1$--$10\,$s$^{-1}$, permeation at nodes and paranodes
\citep{Nilsson2013}), whereas wall \emph{encounters} recur on the perpendicular diffusion
correlation time ($\sim$$0.1$--$1\,$ms): a spin bounces two to three orders of magnitude more
often than it permeates, which is what makes isolating surface relaxivity with $\kappa=0$
physically justified.

\paragraph{Biological idealisations.} Four substrate assumptions bound the biological
reading. First, the effect is carried by the sub-micron calibre tail (the crossing is at
$d^{\ast}\!\approx\!0.17\,\mu$m at the cited $\rho$), so the magnitude depends on how much
sub-micron water the tissue holds --- which we anchor with rat spinal-cord cross-sections
\citep{Zaimi2018}, a coarser-calibre tissue than the human cortical and callosal white matter
that is the clinical target. Human callosal fibres run finer, with a substantial sub-micron
population \citep{Aboitiz1992}, so if anything the rat-cord anchor \emph{under}-states the
tissue-borne effect; a human-EM calibre census is the proper anchor and we do not claim one.
Second, we hold the g-ratio fixed at $0.70$ across calibre, whereas fine axons are relatively
more thickly myelinated (lower $g$) \citep{Stikov2015}; since the crossover
$\mathrm{VF}^{\ast}=1/(1+g)$ depends on $g$, the single sign law is a calibre-averaged
statement, and a realistic $g(d)$ would spread the crossover across calibres rather than
abolish it. Third, we assign the inner and outer walls one relaxivity $\rho_{\mathrm i}=\rho_{\mathrm e}$;
these are chemically different membranes whose relaxivities may differ, a \emph{structural}
(not merely scale) uncertainty that could shift the interior/exterior balance and hence the
$f_{\mathrm{intra}}$ sign, and $\rho$ itself is imported from a histology-fit model coefficient
\citep{Barakovic2023} that may partly absorb exchange or susceptibility. Fourth, the
longitudinal cancellation assumes lumen-preserving (pure g-ratio) demyelination; real MS and
Wallerian degeneration co-vary axon calibre through swelling, beading, atrophy and outright
axonal loss, which alter the very inner-diameter distribution the offset is set by --- so the
clean within-subject cancellation is an idealised bound, and where pathology shifts the calibre
spectrum the cross-region confound re-enters the longitudinal comparison.

\paragraph{Limitations.} We assumed ideal, instantaneous refocusing, isolating the
bulk-versus-surface question; finite-pulse and susceptibility effects are real but
orthogonal and treated elsewhere. This sets the effect in perspective: the
flip-angle/stimulated-echo and fibre-orientation/susceptibility \citep{Birkl2021}
systematics the MWF field already corrects for are themselves of order a percentage point
or more --- comparable to or larger than the surface bias at the cited $\rho$, which is a
further, hitherto-unnamed term of the same family. The MWF magnitudes come from a forward
model carrying the validated rate over a calibre-distributed intra pool
(Sec.~\ref{sec:methods:mwf}), so the absolute numbers inherit its two-/three-pool,
non-exchanging, canonical-Gamma idealisations --- which modulate the magnitude but not the
existence or sign of the term. The robust core needs no estimator: the closed-form rate and
its wall-counting Monte-Carlo validation (a motional-narrowing limit and a geometry-only
simulation with no relaxivity model) agree, and the short-$T_2$ crossing they imply is a
grid-free property of the signal. The claim we stand behind is the inseparability itself,
the sign of its consequence (fine reads myelin-richer), and the order of magnitude that
follows once $\rho$ is fixed.

\section{Conclusion}
\label{sec:conclusion}

The transverse rate underlying microstructure imaging is an intrinsic bulk rate plus a
surface-relaxivity rate $\rho\,(S/V)$ --- both isotropic and TE-linear, hence
non-identifiable, with $S/V$ unresolved. It is the relaxometric face of the wall collisions
diffusion MRI reads as time-dependent extra-axonal diffusion, and it acts with no gradient;
a wall-counting Monte Carlo reproduces it from first principles. Because the interior and
exterior walls carry different $S/V$, the compartments relax differently, and any estimate
that normalises by a TE-weighted $b{=}0$ is biased. On the diffusion side this over-weights
the intra-axonal signal (the leading-order intra-axonal fraction bias) on physiologically
dense white matter --- by ${\approx}12\%$ over the robust packing band at clinical echo
time and the cited $\rho$, with a sign set by a single packing ratio
$(1-\mathrm{VF})/(g\,\mathrm{VF})$ that crosses unity at $\mathrm{VF}^{\ast}=1/(1+g)$, and a
testable, packing-dependent TE drift --- a first-order, systematic bias on a headline metric
that shifts regional and cohort means rather than averaging out. The same physics reads through relaxometry as a smaller, structural
myelin-water bias: the thinnest axons' water is dragged below the myelin window and counted
as myelin, so fine white matter reads myelin-\emph{richer} at equal myelin content
(${\sim}0.33$\,pp, ${\sim}0.82$\,pp of inferred myelin after the $1/W_M$ amplification,
climbing super-linearly at the upper end of the order-of-magnitude-uncertain $\rho$); being
lumen-set, it \emph{cancels} in the within-subject longitudinal change that primary
demyelination traces. Diffusion is in principle the orthogonal axis that could constrain
the geometry directly --- the same wall collisions, read with a gradient --- but realising
it is gradient-limited, left to future work.

\appendix
\section{Surface relaxation as a boundary-local-time estimator}
\label{sec:app:estimator}

The wall-counting of Sec.~\ref{sec:methods:mc} is a Monte-Carlo realisation of the
Brownstein--Tarr Robin boundary condition $D\,\partial_n M = -\rho\,M$
\citep{Brownstein1979}. That boundary condition is not a per-collision penalty: it is a
continuous absorbing wall, and its stochastic representation is the \emph{boundary local
time} $\mathcal L_{\partial\Omega}$ of reflected Brownian motion --- the (measure-theoretic)
contact the walker accumulates at the surface --- through the Feynman--Kac survival weight
$\exp(-\tfrac{\rho}{D}\,\mathcal L_{\partial\Omega})$ \citep{Grebenkov2007}. Any faithful
simulator therefore only has to \emph{estimate} $\mathcal L_{\partial\Omega}$ from the discrete
walk; different estimators agree in the $\Delta t\!\to\!0$ limit but differ at finite step.

We use the \emph{realised-overshoot} estimator: a step that crosses the wall contributes
$2\,d_\perp$ to $\mathcal L_{\partial\Omega}$, with $d_\perp$ the perpendicular overshoot of
that step, so the transverse magnetisation is scaled by $\exp(-2\rho\,d_\perp/D)$ per
encounter --- a \emph{pathwise} estimator weighted by the actual penetration depth. An
alternative, used by the multi-modal simulator \textsc{MCMRSimulator} (v1.0.0) of
\citet{Cottaar2025}, assigns every detected collision the \emph{same} cost
$\exp(-x\sqrt{\Delta t})$ (\texttt{src/evolve.jl}, lines 566--570), i.e.\ it replaces the per-step
penetration with its mean; its documented material rate is $x=\rho\sqrt{\pi/D}$, which we use
as given (no calibration). Both are legitimate; they differ only in whether the collision cost
carries the realised depth or a fixed $\sqrt{\Delta t}$.

\begin{figure}[t]
  \centering
  \includegraphics[width=\linewidth]{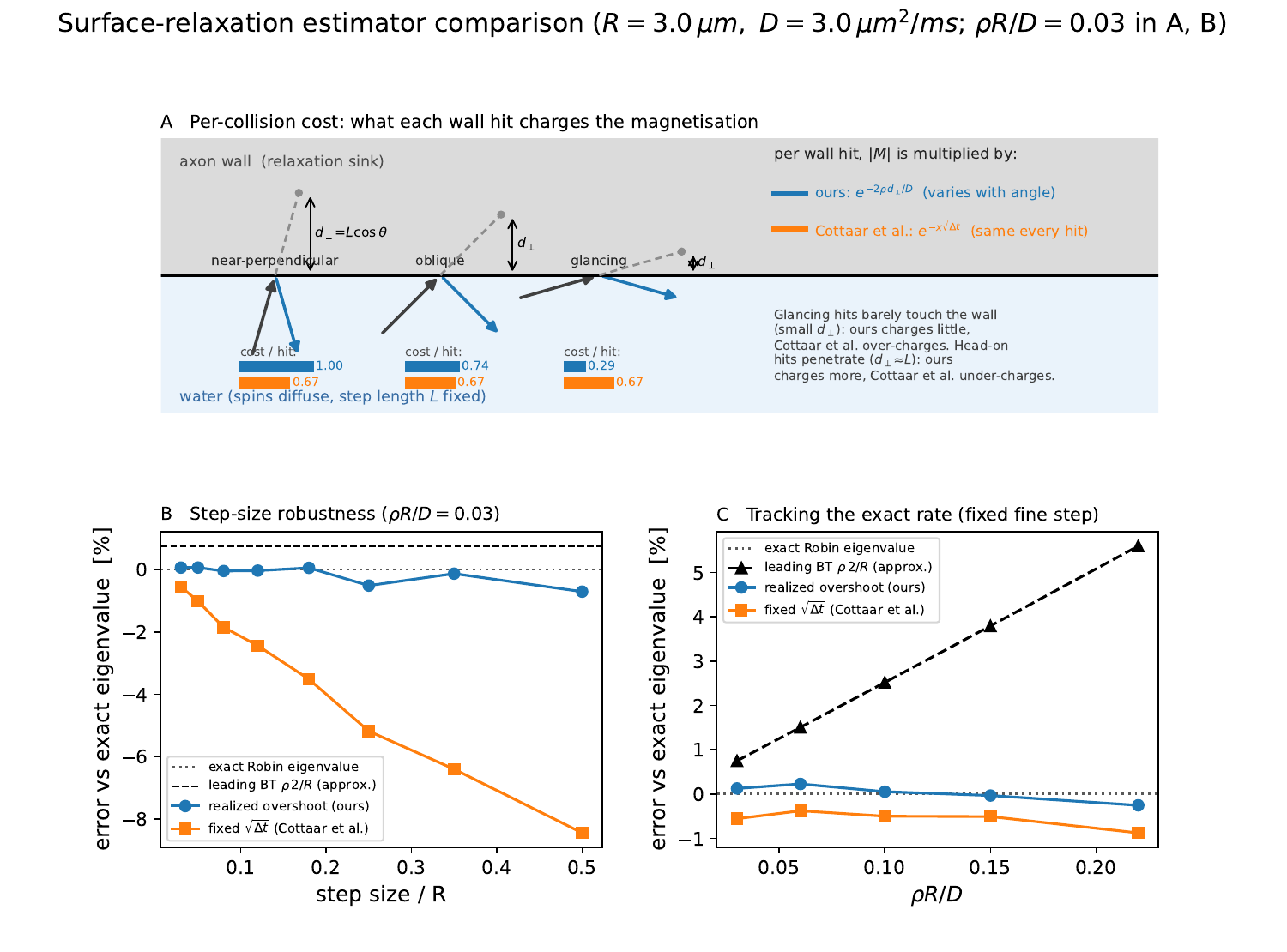}
  \caption{Two Monte-Carlo estimators of the same Brownstein--Tarr surface relaxation, on
  identical collision events of one reflecting walk. \textbf{(A)} Per wall hit: for a fixed step
  $L$ the perpendicular overshoot $d_\perp=L\cos\theta$ grows from glancing to head-on, so the
  realised-overshoot rule charges $e^{-2\rho d_\perp/D}$ (blue, $1.00/0.74/0.29$) while the
  fixed-$\sqrt{\Delta t}$ rule charges the same each hit (orange, $0.67$) --- its mean over-charges
  glancing and under-charges head-on hits. \textbf{(B,C)} Error vs the exact Robin eigenvalue
  (dotted; leading $\rho(2/R)$ dashed): the parameter-free realised-overshoot estimator (ours)
  stays on the exact value across step size (B) and $\rho R/D$ (C); the fixed-$\sqrt{\Delta t}$
  rule (\citealp{Cottaar2025}) drifts to $-8\%$ at a coarse step (B); and the leading-order
  \emph{analytic} formula --- not the estimators --- diverges to $+5.6\%$ as $\rho R/D$ grows (C).}
  \label{fig:app:estimator}
\end{figure}

To isolate the \emph{estimator} --- rather than two codebases, which would confound the
per-collision rule with stepping, detection and RNG --- we run one reflecting walk in a cylinder
cross-section ($R=3\,\mu$m, $D=3\,\mu$m$^2$/ms) and accumulate \emph{both} penalties on the
\emph{same} collisions, referenced to the exact lowest Robin eigenvalue (the root of
$(kR)J_1(kR)/J_0(kR)=\rho R/D$, of which the textbook $\rho\,(2/R)$ is only the
$\rho R/D\!\to\!0$ limit, larger by $\rho R/4D$). This is an \emph{estimator} comparison inside
one code, not a two-codebase cross-validation: reimplementing the fixed-$\sqrt{\Delta t}$ rule on
our own walk holds the stepping, detection and RNG common, so the only independent reference is
the exact Robin eigenvalue --- and it is against that eigenvalue, not against either simulator,
that correctness is judged. Both schemes reproduce it at a fine step
(Fig.~\ref{fig:app:estimator}B,C), confirming our surface term is implemented correctly; the
$\sqrt{\Delta t}$ rule being MCMRSimulator's documented choice, this also shows the two are
consistent in the fine-step limit, but it is not an agreement between two independent codes. We
adopt the realised-overshoot form (the factor $2$ is the local-time normalisation, nothing
fitted) because it is parameter-free and holds at coarser steps, where the fixed-$\sqrt{\Delta t}$
rule drifts to $-8\%$.

The estimator's relative error is a function of the two dimensionless groups $\rho R/D$ and
step$/R$ alone, so the accuracy shown at $R=3\,\mu$m transfers to any calibre: holding both
groups fixed and sweeping the radius from $3$ down to $0.1\,\mu$m leaves the error flat at
$-0.01\%$ (a scale-invariance check in the same script). This is what licenses carrying the
interior rate to the sub-micron calibres that dominate the bias --- they are not an
extrapolation but the \emph{same} dimensionless regime, and in fact an easier one: at the
physical $\rho$ a sub-micron axon has $\rho R/D\!\sim\!10^{-5}$, far deeper in fast diffusion
than the benchmark, where the estimator (and the Brownstein--Tarr rate itself) is most accurate
(panel~C).

\section*{Data and Code Availability}
In keeping with the openness aims of the \texttt{dmrai-lab} project, all materials --- the
manuscript source, the figures, and the Python scripts that regenerate every figure from
scratch (each with its committed numeric data) --- are openly available at
\href{https://github.com/dmrai-lab/dmipy}{\texttt{github.com/dmrai-lab/dmipy}}. The figures
are built on the dmipy computational ecosystem: \texttt{dmipy-fit}, a subsumption and
expansion of the original \texttt{dmipy} toolbox \citep{Fick2019} (the analytical
multi-compartment signal models and the NNLS myelin-water estimator), and
\texttt{dmipy-sim}, a GPU/JAX-accelerated toolbox for physics-complete Monte-Carlo spin
simulation under arbitrary free waveforms, sharing one pulse-sequence and substrate
interface with \texttt{dmipy-fit}. The real cross-sections are the AxonDeepSeg SEM
rat dataset \citep{Zaimi2018}.

\section*{Acknowledgements}
The author thanks the dmrai-lab open-source diffusion- and quantitative-MRI
simulation community.

\bibliography{refs}

\end{document}